\documentclass[aps,prb,twocolumn,groupedaddress,floatfix]{revtex4}

\usepackage{graphicx}
\usepackage{amsmath}
\usepackage[usenames]{color}

\newcommand{\be}{\begin{equation}}
\newcommand{\ee}{\end{equation}}
\newcommand{\bea}{\begin{eqnarray}}
\newcommand{\eea}{\end{eqnarray}}
\newcommand{\bd}{\begin{displaymath}}
\newcommand{\ed}{\end{displaymath}}
\newcommand{\bk}{\mathbf{k}}
\newcommand{\bR}{\mathbf{R}}

\newcommand{\kB}{k_{\mathbf{B}}}

\begin{document}


\title{Realistic heterointerface model for excitonic states in growth-interrupted quantum wells}


\author{Vincenzo~Savona}
\affiliation{Institut de Th\'eorie des Ph\'enom\`enes Physiques, Ecole
Polytechnique F\'ed\'erale de Lausanne (EPFL), CH-1015 Lausanne, Switzerland}

\author{Wolfgang Langbein}
\affiliation{Department of Physics and Astronomy, Cardiff University, Cardiff
CF24 3YB, United Kingdom}

\date{\today}

\begin{abstract}
We present a model for the disorder of the heterointerfaces in GaAs
quantum wells including long-range components like
monolayer island formation induced by the surface diffusion during
the epitaxial growth process. Taking into account both interfaces, a
disorder potential for the exciton motion in the quantum well plane
is derived. The excitonic optical properties are calculated using
either a time-propagation of the excitonic polarization with a
phenomenological dephasing, or a full exciton eigenstate model
including microscopic radiative decay and phonon scattering rates.
While the results of the two methods are generally similar, the
eigenstate model does predict a distribution of dephasing rates and
a somewhat modified spectral response. Comparing the results with
measured absorption and resonant Rayleigh scattering in GaAs/AlAs
quantum wells subjected to growth interrupts, their specific
disorder parameters like correlation lengths and interface flatness
are determined. We find that the long-range disorder in the two
heterointerfaces is highly correlated, having rather similar average
in-plane correlation lengths of about 60 and 90\,nm. The
distribution of dephasing rates observed in the experiment is in
agreement with the results of the eigenstate model. Finally, we
simulate highly spatially resolved optical experiments resolving
individual exciton states in the deduced interface structure.
\end{abstract}



\maketitle

\section{Introduction}
The interfaces of a semiconductor quantum well (QW) always present some amount of disorder as a natural consequence of the physics of the growth process. This
disorder determines to a large extent the excitonic optical response both in frequency and time domain. Optical spectroscopy on QW excitons is therefore able
to probe the interface disorder. However, the link between the optical response, particularly resonant Rayleigh scattering (RRS), and the details of the
interface structure, is not straightforward and calls for a microscopic quantum model. QW interface fluctuations affect both the potential and the kinetic
contributions to the total exciton center-of-mass (COM) energy, resulting in a Schr\"odinger equation for the exciton COM wavefunction along the QW plane,
including a static disorder potential that is related to the actual interface structure. The exciton COM wavefunction is proportional to the exciton
macroscopic interband polarization and relates therefore directly to the linear optical response of the system. The eigen-states in this model present both a
spatial localization within the QW plane and an inhomogeneous distribution of eigen-energies. When QWs are grown by molecular beam epitaxy, interface disorder
is mostly determined by surface diffusion and segregation. When using semiconductor alloys for the well or the barrier, also alloy disorder becomes important.
While surface diffusion tends to create laterally extended regions with a well-defined monolayer (ML) thickness, segregation and alloy disorder result in
nanoroughness on an atomic length scale, smaller than the exciton Bohr radius \cite{WarwickAPL92}. The interruption of the epitaxial growth at the QW
interfaces leads to a longer surface diffusion length, creating ML islands in the 100\,nm size range that have been observed in various experimental
studies\cite{BimbergJVST92,KopfJAP93,BernatzJAP99}. When ML islands are sufficiently\cite{CastellaPRB98} large compared to the excitonic Bohr radius, the
exciton spectrum can split into several peaks of reduced inhomogeneous broadening\cite{GammonPRL91}. In a simple picture these peaks are resulting from exciton
states present within flat regions of different integer ML thickness, while the inhomogeneous broadening results from a residual nanoroughness. To create from
this picture a realistic description of the ensemble properties of the exciton in the QW, one needs to consider the following issues: (i) the statistical
distribution of size and shape of regions with different ML thickness of the interfaces; (ii) the difference of the structural parameters and the statistical
correlation of the two interfaces; (iii) the lateral quantum confinement energy of the exciton states; (iv) the coupling of the exciton states to light. Once
the statistical properties of the interface structure are described by a model with a small number of parameters, these parameters can be deduced from the
measured RRS spectrum and dynamics.

Existing theoretical models of RRS in QWs were often based on very simplified
assumptions for the shape of the disorder potential. The simplest picture,
introduced in the original work on RRS theory by Savona and
Zimmermann\cite{SavonaPRB99} consisted in assuming a random potential with
Gaussian or exponential correlation length in space. Although this approach can
explain the principial properties of RRS,\cite{KocherscheidtPRB02} it is unable
to predict the dynamics of more complex situations like the one considered
here. In a previous work \cite{KocherscheidtPRB03}, we had developed a
rudimentary extension of the disorder model including a short- and a long-range
disorder component, for the modeling of RRS in samples where one of the two
heterointerfaces was growth interrupted.

In this work we apply the microscopic theory of the RRS optical response of
excitons \cite{SavonaPRB99} to a situation where a pronounced ML splitting is
present. To this purpose, we develop a comprehensive model of the disorder
potential based on a separate description of the direct and inverse
heterointerfaces of the QW. The model accounts for a ML island structure at
both QW interfaces, with different correlation lengths, anisotropy factors and
statistical distributions of the integer values of the ML thickness. It also
includes nanoroughness and a parameter that expresses the statistical
correlation between the two interfaces.
The experimental RRS data, both spectrally and time-resolved, were taken
on a GaAs/AlAs QW sample grown with long growth interruptions at both
interfaces. The amount of deposited GaAs is varying slowly across
the lateral position on the sample, so that a fine control of the
{\em average} QW thickness by a small fraction of one ML is achieved.
This in turn produces a continuous evolution of
the ML-split exciton spectrum, which replicates itself after a
thickness change of an integer ML with only an energy shift given by
the corresponding difference in QW confinement energy.
Correspondingly, also the time-resolved RRS shows a peculiar
dependence on the fractional ML thickness. We use this rich
phenomenology to determine the parameters of a detailed interface
model that we develop. The calculated RRS response is confronted with the
measured spectral and time-dependent RRS signatures, and the effects
of the various interface properties are elucidated. This analysis
allows an estimation of the actual parameters characterizing the
investigated sample -- to a much larger extent than in previous
studies based on the same RRS theory. Specifically we find an almost
perfect correlation between the direct and inverse QW interfaces --
a novel result that is of general relevance for epitaxial QW
samples. In the second part of the work, we turn to the numerical
computation of the actual exciton COM eigenstates as resulting from
the disorder model. This part of our study is required in order to
account for state-dependent radiative recombination and phonon
scattering rates, which prove important especially in the RRS
dynamics. It provides us with further insights into the microscopic
structure of the single exciton states and their optical properties.

The paper is organized as follows. In Section II we describe the
time-propagation RRS theory and the interface disorder model, followed by a
study of the dependence of the simulated RRS on the model parameters. Section
III describes the experimental data and the features of the QW interfaces, as
deduced from the comparison to the model calculations. Section IV is devoted to a
microscopic simulation of the exciton eigenstates and of their radiative and
phonon scattering rates. Results for ensemble averages and individual disorder
realizations are presented. Section V contains the conclusions and outlook of
the work.

\section{Theoretical description}
The present model is based on the theory of RRS within the exciton COM approximation, as derived by Savona and Zimmermann \cite{SavonaPRB99}. The exciton
interband polarization $P$ is proportional to the exciton COM wavefunction along the QW plane and therefore evolves according to the time-dependent
Schr\"odinger equation
\begin{eqnarray}
-i\hbar\frac{\partial P({\bf R},t)}{\partial t}&=&\left(-\frac{\hbar^2}{2M}\nabla^2+V({\bf R})\right)P({\bf R},t)\nonumber\\
&+&\mu E_{\rm in}({\bf R},t)\,, \label{RRS}
\end{eqnarray}
where ${\bf R}=(x,y)$ is the position in the QW plane, $M$ is the in-plane exciton mass, $\mu$ is the dipole matrix element of the excitonic transition, and
$E_{\rm in}({\bf R},t)$ is the resonant excitation field. In the present model, we neglect the vector character of the electric field and of the interband
polarization. Rayleigh scattering is expected to conserve the orientation of the polarization vector of the scattered field, whereas the long-range exchange
part of the Coulomb interaction can induce a rotation of this vector on a timescale of several tens of picoseconds\cite{LangbeinPSS00a,ZimmermannBook03}. The
static COM potential $V({\bf R})$ is an effective quantity, related to the actual exciton energy fluctuations via a convolution with the 1s-exciton relative
wavefunction \cite{ZimmermannBook03}. Differently from the usual definition, we have incorporated here the average 1$s$ exciton energy $E_{\rm X}$ into $V({\bf
R})$. We detect the RRS in the far field and neglect time delays due to light-propagation. In linear response theory, the RRS amplitude $E_{\rm out}(\bk,t)$ is
proportional to the Green's function $G(\bk,\bk_{\rm in},t)$ of the Schr\"odinger equation (\ref{RRS}), expressed in momentum
space,\cite{SavonaPRB99,ZimmermannBook03} where $\bk_{\rm in}$ is the momentum of the incident electromagnetic field $E_{\rm in}(\bk_{\rm in},t)$, assumed to
be a plane wave. To determine the spectrally resolved RRS $E_{\rm out}(\bk,\omega) = \int_0^\infty E_{\rm out}(\bk,t)e^{(i\omega-\gamma)t} dt$, we introduce a
phenomenological polarization decay rate $\gamma$. The resulting RRS intensity
\be I_{\rm RRS}(\bk,t) \propto \left| E_{\rm out}(\bk,t)\right|^2, \quad I_{\rm
RRS}(\bk,\omega) \propto \left| E_{\rm out}(\bk,\omega)\right|^2 \ee
is averaged over a large statistical ensemble of disorder realizations. In a similar way, we can define the exciton COM spectral function
$A(\bk,\omega)=-2\mbox{Im}\{G(\bk,\bk,\omega)\}$ as proportional to the imaginary part of the COM Green's function at equal momentum arguments. The spectral
function at momentum $\bk_{\rm in}$, averaged over disorder realizations, is proportional to the exciton absorption spectrum  (the same average is identically
zero if the Green's function is taken with two differing momentum arguments). The simulated spectral function will be compared, in this and in the next
Section, to the measured excitonic absorption. Throughout the paper, when quoting values for frequencies or rates in energy units, they are multiplied by
$\hbar$.

\subsection{Model for the interface structure}

\begin{figure}
\includegraphics*[width=\columnwidth]{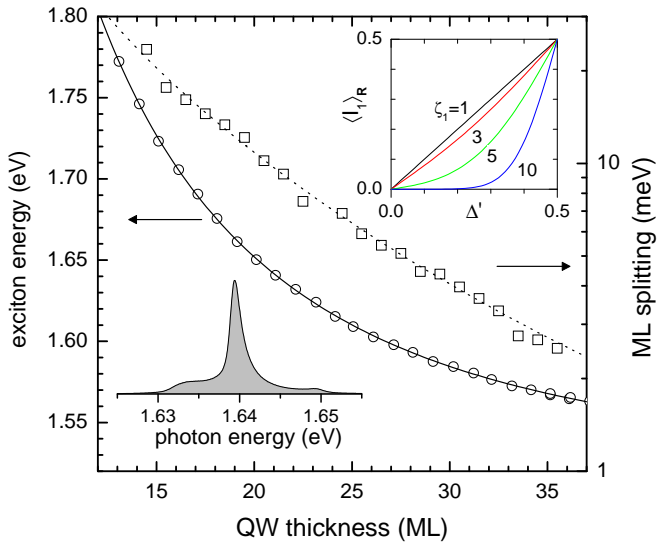}
\caption{\label{fig:Exenfit} Measured exciton transition energy (circles) versus nominal QW thickness $\Delta$, and the fitted $F(\Delta)$ (solid line) using
($a_0=1.516, a_1=-0.233, a_2=92.5, a3=-806, a_4=2739$)\,eV. The measured exciton PL peak energy at a thickness resulting in a dominating single ML peak (see
lower inset) is attributed to the exciton potential $F(\Delta)$ at +0.15 fractional ML thickness. This slight shift is needed to account for the quantum
confinement due to short-range disorder, which is shifting the peak in the optical spectrum to slightly lower energy with respect to the nominal zero-point
(spatial average) of the disorder potential (see e.g. Fig. 5 in Ref. \onlinecite{ZimmermannBook03}). The value +0.15 ML is giving the best fit between RRS
measurements and simulations at $\Delta \approx 38$. Additionally, the measured ML splitting (squares) and its fit $dF/d\Delta$ (dashed line) is shown. The
upper inset shows the average ML coverage of the first interface $\langle I_1 \rangle_\bR$ as function of $\Delta'$. }
\end{figure}

In order to model the exciton COM potential in presence of ML islands in the heterointerfaces, as in the QWs studied here, the following procedure is used.
First, a random white-noise function $w_\alpha({\bf R})$ is generated, such that $\langle w_\alpha({\bf R})w_\alpha({\bf R}^\prime)\rangle=\delta({\bf R}-{\bf
R}^\prime)$. The index $\alpha=1,2$ denotes the two QW heterointerfaces. We then derive a spatially correlated function of unitary standard deviation and zero
mean value
\be W_\alpha  \left( {\bR} \right) \propto \int {d{\bf{R'}}\,w_\alpha\left(
{{\bf{R'}}} \right)\exp \left( { - \frac{{\left( x-x^\prime \right)^2
}}{{2\xi_{x ,\alpha}^2 }}- \frac{{\left( y-y^\prime \right)^2 }}{{2\xi_{y
,\alpha}^2 }}} \right)} \label{W12} \ee
Here we have introduced the Gaussian correlation lengths $\xi_{j,\alpha}$, with $j=x,y$, which will determine the length scale of the ML islands along the $x$
and $y$ directions, respectively. The separate length scales for the two directions allow to describe a spatial anisotropy of the ML structure. We assume equal
orientation of the anisotropy axis for both interfaces since we expect the microscopic origin of the anisotropy to be given by the anisotropic reconstruction
of the [001] growth surface along the [110] in-plane direction, equal for the whole epitaxially grown structure. We express the two correlation lengths as a
function of an anisotropy parameter $\epsilon_\alpha$ and of an isotropic correlation length $\xi_\alpha$, defining
$\xi_{x,\alpha}=\epsilon_\alpha^{1/2}\xi_\alpha$ and $\xi_{y,\alpha}=\epsilon_\alpha^{-1/2}\xi_\alpha$. We proceed by generating a ML step function from
$W_\alpha$ that models the property of a growth surface to form compact regions of integer ML coverage due to the lateral binding energy \cite{HeynPRB97} of
the adatoms, as opposed to a random arrangement. We use the function
\be I_\alpha \left( {\bR} \right) = {\rm{Round}}\left[ {\frac{{W_\alpha \left(
{\bR} \right)}}{{\zeta_\alpha }} + \Delta^\prime +
\Delta\cdot\delta_{\alpha,2}} \right]\, , \label{I12} \ee
where the function Round[$x$] denotes the nearest integer of a real number $x$. The quantity $\zeta_\alpha$ is the interface {\em flatness}, controlling the
height variance of the interface to be approximately $\zeta_\alpha^{-1}$ MLs. The parameter $\Delta^\prime \in \left[ {0,1} \right]$ is a fractional ML shift
that describes the fractional ML coverage of the first interface, i.e. the AlAs surface on which the GaAs QW is grown. The spatial average of the ML coverage
$\langle I_1 \rangle_\bR$ is approximately $\Delta'$ for interfaces that consist of several ML steps, i.e. $\zeta \lesssim 3$. This is the case for the
simulations presented later on. For flatter interfaces the fractional ML coverage is a non-linear function of $\Delta'$ (see upper inset of
Fig.\,\ref{fig:Exenfit}). The same arguments hold for $\langle I_2 \rangle_\bR$ and $\Delta'+\Delta$. Due to an unavoidable slight misorientation ($>$1'') of
the substrate with respect to the [001] direction, the fractional ML coverage of the interface is expected to vary over many ML thicknesses across the excited
surface area, which has a lateral extension of about $10^6$ ML (300\,$\mu$m). We will therefore use a uniform distribution of $\Delta^\prime$ over $\left[
{0,1} \right]$ in the statistical ensembles for the simulations. The quantity $\Delta$ is shifting the second heterointerface with respect to the first one, as
specified by the Kronecker $\delta_{\alpha,2}$ in Eq.\,(\ref{I12}). It defines the average distance between the first and the second heterointerface, i.e. the
average QW thickness. This thickness can be accurately varied in the experiment by displacing the laser spot along the QW plane since the average QW thickness
is given by the deposited amount of GaAs within the surface diffusion length and varies only by about 0.1\,ML over 300\,$\mu$m lateral extension
\cite{LeossonPRB00}. In this argument, we assume that Ga and Al atoms do not evaporate from the surface once deposited. This assumption is justified for the
growth parameters used for the investigated samples.

We continue by deriving a local exciton energy $U\left( {\bR}
\right) = F\left( {I_2 \left( {\bR} \right) - I_1 \left( {\bR}
\right)} \right)$ from the local ML-thickness $I_2- I_1$. The
dependence of the exciton energy on the ML-thickness has the form
\be F(\Delta)=a_0+a_1/\Delta+a_2/\Delta^2+a_3/\Delta^3+a_4/\Delta^4 \ee
where the parameters $a_n$ are fitted to the experimental data
\cite{LeossonPRB00} obtained on the same sample by measuring the exciton peak
position at equivalent fractional ML thicknesses over a large range of
thicknesses (see Fig.\,\ref{fig:Exenfit}). A random, spatially uncorrelated
potential $u\left( {\bR} \right)$ is summed in order to account for short-range
disorder originating from segregation. The COM potential $V(\bR)$ is obtained
by convoluting $U(\bR)+u(\bR)$ with the electron and hole probability
distribution in the 1s-exciton\footnote{The electron and hole wavefunctions in
the in-plane coordinate ${\bf R}'$ are $\propto \exp(-M/m_{e,h}|{\bf R}- {\bf
R}'|/a_{\rm B})$ with the in-plane masses $m_{e,h}$ of electrons and holes. For
$m_e=m_h$, the relationship used in our work is obtained. In other cases, the
convolution function should contain a sum of the electron and hole probability,
weighted with the respective interface potential (see
Ref.\,\onlinecite{ZimmermannBook03}). Within the rigid exciton approximation,
the difference to the chosen approach is not relevant.} $|\phi({\bf
R}')|^2\propto\exp(-|{\bf R}-{\bf R}'|/a_{\rm B})$, using an exciton Bohr
radius $a_{\rm B}=8\,\mbox{nm}$. This value was deduced from the measured
diamagnetic shift $\alpha_d=33\,\mu$eV/K and the reduced mass
$\mu^*=0.0586\,m_0$ for a 37ML thick GaAs/AlAs quantum well \cite{SmithPRB89}
using the formula $\alpha_d=a_{\rm B}^2(3e^2)/(16\mu^*)$ valid for an in-plane
exciton wavefunction proportional to $\exp(-r/a_{\rm B})$, with the in-plane
electron-hole distance $r$. The amplitude of the short-range disorder $u({\bf
R})$ is parameterized by $\sigma$, expressing its standard deviation after the
convolution with the 1s-exciton relative wavefunction.

Within this model, particular attention must be payed to the statistical correlation between the ML fluctuations of the two interfaces. In the epitaxial
growth, a substantial amount of correlation can be expected between the ML fluctuations of the lower and upper interfaces \cite{PonomarevAPL04}. We introduce
this correlation into our model via the two white-noise functions $w_\alpha({\bf R})$ that enter Eq.\,(\ref{W12}). The limiting case of fully correlated
heterointerfaces is obtained by taking $w_1=w_2$. Conversely, the fully uncorrelated case results from using two uncorrelated white noise functions $w_1$ and
$w_2$. In order to model the intermediate case of partial correlation, we introduce two mutually uncorrelated white-noise functions $w({\bf R})$ and
$w^\prime({\bf R})$. We could then proceed by simply defining $w_1=w$ and $w_2=\kappa w+(1-\kappa^2)^{1/2}w^\prime$, therefore assuming a uniform correlation
$\kappa$ for all the Fourier components of $w_{1,2}$. However, the total number of Ga atoms deposited over length scales larger than the surface diffusion
length has to be conserved in the growth process. Even though the atoms are deposited at random in molecular beam epitaxy, the total number of Ga atoms
contained in the QW within an area given by the surface diffusion length is $>10^6$, so that the statistical error in the QW thickness is below $10^{-3}$, a
fluctuation that can be neglected compared to the interface roughness. The long-wavelength components of the ML fluctuations in the two interfaces have
therefore to be fully correlated. In order to accommodate this requirement, we modify the above approach for the long-wavelength component of $w_2$. Using the
Fourier transforms $w_\bk$ and $w^\prime_\bk$ of the two generating functions, we define the function $w_2({\bf R})$ as the inverse Fourier transform of
\begin{eqnarray}
w_{2,\bk}=f_kw_\bk+\left(1-f_k^2\right)^{1/2}w^\prime_\bk \label{wcorr}
\end{eqnarray}
with \be f_k=(1-\kappa)\exp\left(-\frac{k^2}{l_c^2}\right)+\kappa\,, \ee where
$l_c$ is a surface diffusion length. In this way, the parameter
$\kappa\in[0,1]$ still represents the correlation between the two interfaces
for the Fourier components having wavelengths shorter than $l_c$, while for
longer wavelengths the interfaces are fully correlated as dictated by the
conservation of Ga atoms. It the following simulations, $l_c$ is always much
larger than the correlation lengths $\xi_{\alpha}$, and has therefore no
significant effect. We used $l_c=500$\,nm.

The conservation of Ga atoms also requires that the long wavelength components
of $W_{1,2}$ created from $w_{1,2}$ are equal, which relates the correlation
length and the flatness parameters of both interfaces by
$\xi_1/\zeta_1=\xi_2/\zeta_2$. We will assume this relation to hold throughout
this work.

\begin{figure}[t]
\includegraphics*[width=0.98 \columnwidth]{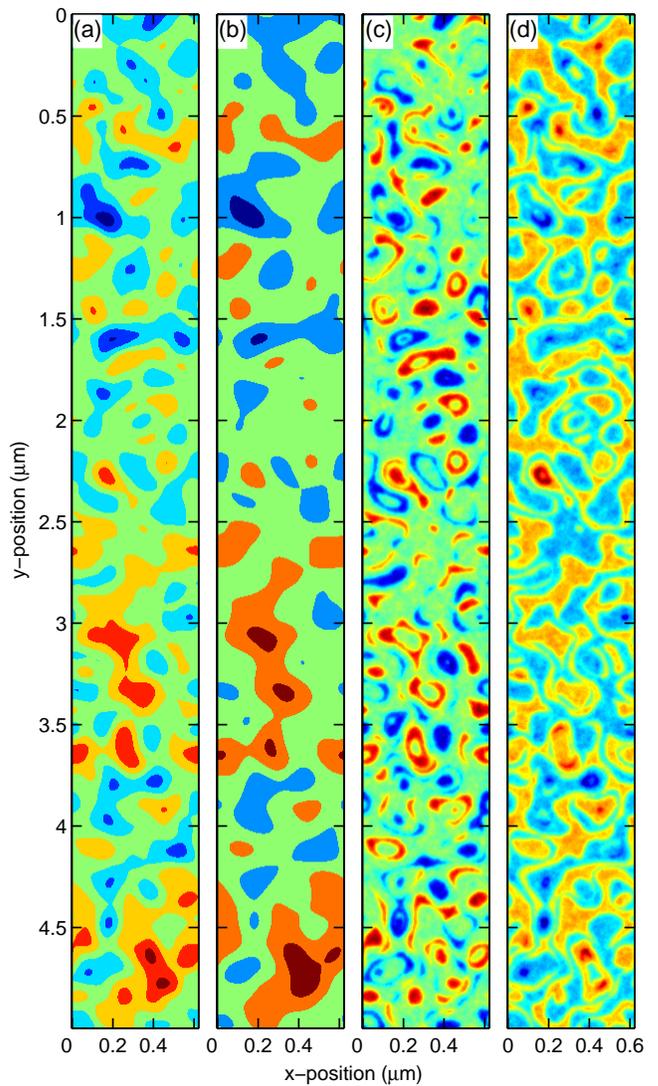}
\caption{\label{fig:potential}(a) Color plot of the first interface profile $I_1({\bf R})$ for $\Delta=38.0$ while $\Delta^\prime$ is varied in the interval
$[0,1]$ along the y-position. The color steps represent ML height steps. (b) Same for the second interface profile $I_2({\bf R})$. (c) COM potential $V(\bR)$
for an integer monolayer thickness $\Delta=38.0$. (d) Same for half-integer monolayer thickness $\Delta=38.5$. The parameters for this potential are given in
Table\,\ref{tab:bestfitpar}, and the correlation is $\kappa=1$. The color scale in c) \& d) covers 5\,meV.}
\end{figure}

\subsection{Simulations}

The time-dependent RRS field is simulated by solving the
time-dependent Schr\"odinger equation with the potential $V(\bR)$,
according to Eq.\,(\ref{RRS}). The Fourier transform of the
interband polarization then provides the RRS amplitude in wavevector
and frequency domain \cite{SavonaPRB99}. The RRS intensity computed
in this way is averaged over $10^3-10^4$ realizations of $V(\bR)$ to
create a statistically relevant ensemble average, converging to its
asymptotic value within $3-1$\% accuracy. The remaining statistical
fluctuations are only appreciable in the time-resolved RRS intensity
at long times, at which the asymptotic RRS dynamics is very slow. It
was shown \cite{SavonaPRB99} that the realization average coincides
with the speckle average which is performed in the experiment by
using a large excitation region of about (200\,$\mu$m)$^2$ and
averaging over a small span of scattering directions. \cite{LangbeinPRL99} This
correspondence is very useful, as simulating directly a large area
is numerically prohibitive. We use typically a (500\,nm)$^2$
simulation area over $64 \times 64$ grid points with periodic
boundary conditions. We assume a plane-wave excitation field at
$\bk=0$ and take the scattered RRS field at the smallest available
wavevector value k=$4\pi/\mu$m on the simulation grid. The average
ML thickness $\Delta$ is known for each position on the sample by a
previous study \cite{LeossonPRB00}. The free parameters used in this
model are therefore the correlation lengths $\xi_{\alpha}$, the
flatness $\zeta_1$, the anisotropy constants $\epsilon_\alpha$, the
correlation $\kappa$, and the short-range disorder amplitude
$\sigma$.

In this Section we proceed by discussing the simulated, spectrally or
time-resolved RRS intensity for the disorder parameters that best describe our
experimental data, which are presented in full detail in the next Section (see
Fig.\,\ref{fig:ExpRRS}). These parameters were obtained by a trial-and-error
procedure that consisted in (i) performing a simulation with a given parameter
set, (ii) comparing to the measured data (both spectral and time-resolved), and
(iii) adjusting heuristically the parameters and going back to step (i). A
standard best-fit procedure is made prohibitive by the long computing time
required for a single simulation (a few hours for TP, a few days for ES).
Starting from this parameter set, indicated in Table \ref{tab:bestfitpar}(TP)
we discuss how varying the different parameters of the model affects the
outcome of the simulations, also in light of previous reports concerning the
interface structure in epitaxial crystal growth. This analysis also provides
estimations of the relative accuracy of the parameter values determined from
the comparison to the experiments.

\begin{table}[b]
\begin{tabular}{|l|c|c|c|c|c|c|c|c|}
\hline
Model&$\xi_1$&$\xi_2$&$\zeta_1$&$\zeta_2$&$\epsilon_1$&$\epsilon_2$&$\sigma$&$\gamma$\\
\hline
TP & 65\,nm & 87\,nm & 1.2 & 1.6 & 1.0 & 1.0 & 0.8\,meV & 10\,$\mu$eV\\
ES & 55\,nm & 82\,nm & 1.46 & 2.2 & 1.0 & 1.0 & 0.8\,meV & -\\
\hline
\end{tabular}
\caption{\label{tab:bestfitpar} Parameters of the interface model obtained by adjusting the simulations to the measured RRS data for the time-propagation
(TP) and the eigenstate (ES) calculations.}
\end{table}

Fig.\,\ref{fig:potential}(a) and (b) show examples of the two interface functions $I_1({\bf R})$ and $I_2({\bf R})$ for the interface parameters found to
reproduce the measured RRS (see Table \ref{tab:bestfitpar}:TP). The fractional ML shift $\Delta^\prime$ was varied linearly from 0 to 1 along the $y$
direction. These false color pictures illustrate how the next ML level gradually develops when moving along the $y$ direction. The two interfaces are fully
correlated ($\kappa=1$), as is reflected in the similar shapes of the two ML patterns. The non-equal correlation lengths $\xi_1<\xi_2$ result in smaller
features and more irregular patterns being present on the first interface (a). The resulting COM potential $V({\bf R})$ for integer and half-integer ML
thickness $\Delta$ of the QW are displayed in (c) and (d). Both pictures show that the variation of the fractional ML shift $\Delta^\prime$ is compensated when
subtracting the ML profiles (Eq.\,\ref{I12}) of the two interfaces, and does thus not significantly affect the COM potential. On the other hand, the difference
between integer and half-integer ML thickness appears clearly in the plots. Because of interface correlation, for integer ML thickness the ML steps on the two
interfaces appear synchronously and, in the limiting case of identical interfaces, the QW thickness would remain constant. \footnote{In this case, our model
results in a vanishing disorder potential, as the change in the absolute exciton position in z-direction is not considered. However, realistically such a
disorder contribution would exist, giving rise to repulsive potentials at the ML steps. We expect these contributions to the disorder potential to be weak
compared to the ML thickness fluctuations as long as their lateral extension is larger than the exciton bohr radius.} The fluctuations in the QW thickness are
thus due to the different characteristic sizes of ML islands on the two interfaces which give rise to ring and stripe patterns in the QW thickness and thus in
$V({\bf R})$, as can be seen in Fig.\,\ref{fig:potential}(c). In the case of half-integer ML thickness the steps on the two interfaces are out of phase, and
the ML steps are likely to appear on one interface only, giving rise to flat regions in $V({\bf R})$ as displayed in Fig.\,\ref{fig:potential}(d), which are
representing two rather well defined ML thicknesses.

\begin{figure}
\includegraphics*[width=8cm]{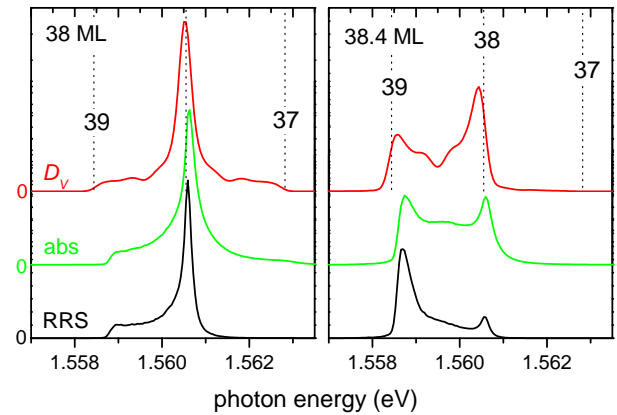}
\caption{\label{fig:Overview} Calculated spectral disorder potential
density {\bf $D_V(\omega)$}, absorption (abs), and RRS intensity
(RRS) for $\Delta=38.0$ and $38.4$ ML, other parameters as in
Fig.\,\ref{fig:potential}. Data are vertically offset for clarity.
The dotted lines represent the potential $F(\Delta)$ for the
labelled integer $\Delta$.}
\end{figure}

The spectral density
\be D_V(\omega)=A^{-1}\int\delta(V(\bR)-\hbar\omega)d\bR \ee
of the resulting disorder potential is given in Fig.\,\ref{fig:Overview}. For integer ML thickness $\Delta=38.0$, $D_V$ has a dominant peak around $F(38)$,
with tails towards the high and low energy extending to the potentials of the adjacent integer ML thicknesses. Due to the averaging over the exciton size, the
regions of adjacent ML thickness do not result in such distinct peaks since they have a small spatial extension. The calculated absorption peak is shifted to
higher energies due to the finite mass of the exciton giving a lateral quantization energy. Furthermore, only the lower energy ML is visible in the absorption,
but not the higher one. This asymmetry relates to the small momentum of the light, which essentially probes the $k=0$ components of the wavefunctions. With
increasing energy, the wavefunctions get more nodes, i.e. a smaller $k=0$ component, so that only the lower part of the potential landscape relates to
optically active exciton states. The RRS spectrum shows a similar behaviour as the absorption, but in a more pronounced way. This is expected because the RRS
intensity -- being proportional to the squared COM Green's function -- varies as the fourth power of the COM wavefunction at small momentum, as opposed to the
second power dependency characterising the absorption spectrum. For about half-integer ML thickness $\Delta=38.4$, $D_V(\omega)$ is distributed between $F(38)$
and $F(39)$, as expected. The absorption and the RRS shows two peaks, with the lower energy peak dominating, again due to the dominant $k=0$ component of the
lowest energy eigenstates.

\begin{figure}
\includegraphics*[width=\columnwidth]{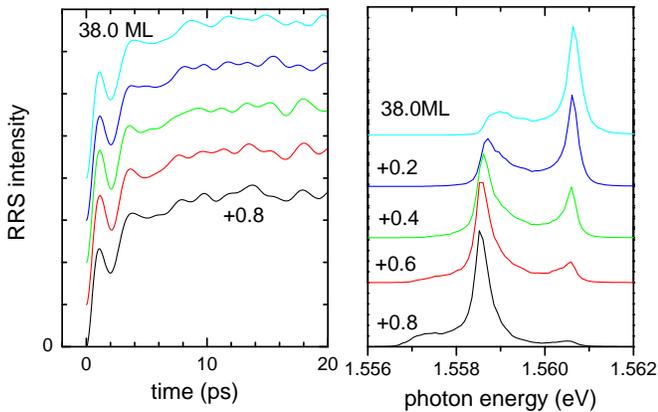}
\caption{\label{fig:RRS38Flat} Calculated time- (a) or spectrally (b) resolved RRS intensity versus the fractional ML thickness $\Delta$ for a GaAs/AlAs
interface of long and a AlAs/GaAs interface of short correlation length ($\xi_1=25$\,nm, $\xi_2=200$\,nm), and a ML flatness $\zeta_1=2.5$. Data are offset for
clarity.}
\end{figure}

\subsubsection{Long correlation length of GaAs/AlAs interface}

The parameters of Table \ref{tab:bestfitpar} define fully correlated interfaces with similar correlation lengths $\xi_{1}=65$\,nm and $\xi_{2}=87$\,nm. From
structural investigations, it could be expected that the islands on the GaAs surface are much larger than on the AlAs surface due to the largely different
surface mobility at the growth temperature of $630^\circ$C. On GaAs surfaces after 60\,s growth interruption, islands of several hundred nanometers were
observed \cite{BernatzJAP99,KopfJAP93,BimbergJVST92}. Earlier investigations, however, measured the GaAs surface after cooling, which can have a significantly
larger correlation length than interior interfaces \cite{BernatzJEM00}. In order to test this expectation, we therefore discuss now the RRS spectra of excitons
in interface structures with a large correlation length of the GaAs to AlAs interface of $\xi_2=200$\,nm. This is much larger than the localization length of a
ML island \cite{LeossonPRB00,CastellaPRB98} $\xi_0=\pi\hbar/\sqrt{2M (dF/d\Delta})\approx 18$\,nm for $\Delta=38$. This interface thus does not significantly
influence the exciton localization. The free parameters are then the AlAs to GaAs interface correlation length $\xi_1$ and flatness $\zeta_1$. The variation of
the RRS spectra with increasing $\Delta$ for 38 to 39 is best reproduced using $\xi_1=25$\,nm and $\zeta_1=2.5$, as displayed in Fig.\,\ref{fig:RRS38Flat}. The
resulting RRS dynamics, however, is essentially independent of $\Delta$, in contrast to the experimental results (see Fig.\,\ref{fig:ExpRRS}). This finding is
insensitive to the specific values of $\xi_1$ and $\zeta_1$, and shows that two interfaces with largely different correlation lengths are not consistent with
the measured RRS dynamics.

\begin{figure}
\includegraphics*[width=\columnwidth]{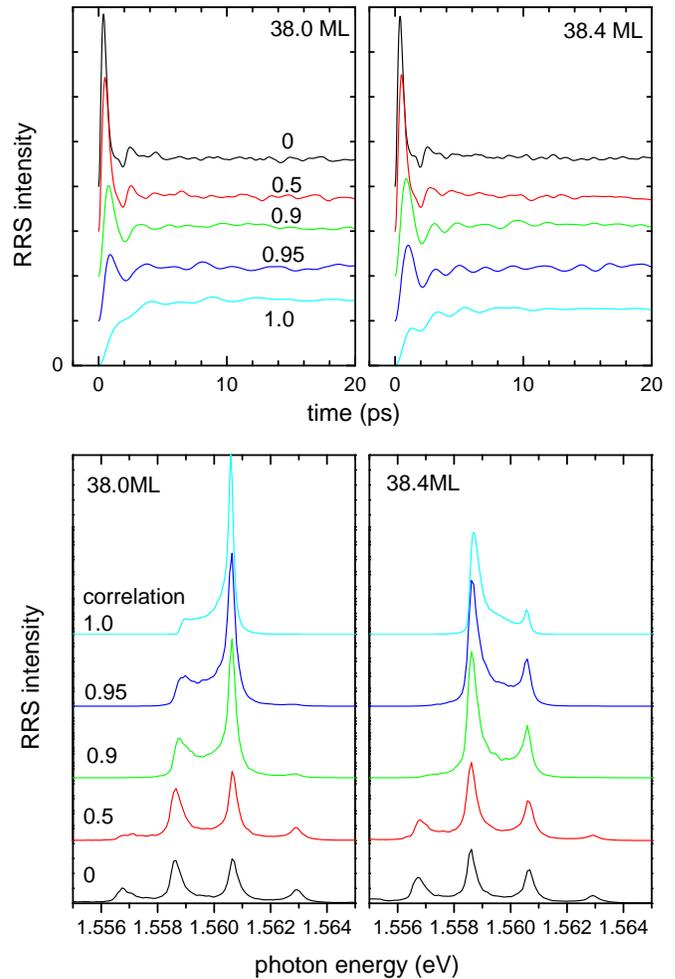}
\caption{\label{fig:RRS38Cor} Calculated time- (a) or spectrally (b) resolved
RRS intensity versus the interface correlation $\kappa$. Other parameters as in
table\,\ref{tab:bestfitpar}. Data are offset for clarity.}
\end{figure}

\subsubsection{Interface Correlation}

For the case of largely different lengths $\xi_{1,2}$ of the two interfaces as discussed in the previous Subsection, the correlation $\kappa$ is not important
since the ML islands of the interfaces are very different. For similar correlation lengths of the two interfaces instead, $\kappa$ is important, since for full
correlation ($\kappa=1$) only the small difference between the two correlated interfaces remains, leaving only narrow flat regions in $V({\bf R})$ as seen in
Fig. \ref{fig:potential}(c). Without correlation ($\kappa=0$), on the other hand, the difference between the uncorrelated interfaces is similar to the case of
a single interface with a reduced flatness. Starting from the parameters in Table \ref{tab:bestfitpar}(TP), we show in Fig.\,\ref{fig:RRS38Cor} the effect of a
decreasing correlation on the RRS spectra and dynamics. Already for a small deviation from full correlation ($\kappa=0.95$), for which the RRS spectrum is not
significantly changed, the RRS dynamics develops an initial peak indicating the presence of a disorder component of large correlation length\cite{SavonaPRB99}.
This component results from the difference between the $k>1/\xi_{2}$ Fourier components of $W_{1}$ and $W_2$, which does not vanish for non-perfect correlation
$\kappa<1$. For even smaller values of $\kappa$, the RRS spectrum gradually develops more ML peaks ($\Delta=37,39$), and the initial peak in the RRS dynamics
gets more pronounced. The RRS dynamics becomes essentially independent of the fractional $\Delta$, a behavior inconsistent with the experiment. We thus
conclude that the ML steps on the two interfaces have to be nearly perfectly correlated. This finding suggests that for our samples the surface diffusion
length $l_c$ is actually similar to the correlation lengths $\xi_\alpha$, so that the almost perfect interface correlation is created by the conservation of
the deposited Ga atoms. We would expect this finding to depend on the duration of the growth interruption at the GaAs to AlAs interface and the growth
temperature. For long growth interruption times and increasing temperature, the surface diffusion length is increasing compared to the island size, as the
surface diffusion distance is increasing with $\sqrt{t}$, while the island size is thermodynamically limited. With increasing duration of the growth
interruption (GI) and growth temperature, the interface correlation is therefore expected to be diminished. This is actually supported by previous experiments.
Yu et al. \cite{YuJAP95} have studied the PL spectra of GaAs/AlAs QWs depending on the GI time. They observe (Fig.\,1 of Ref. \onlinecite{YuJAP95}) the
formation of the ML splitting during the first 60\,s GI, with only 2 ML peaks present, while for 120\,s GI, already 4 peaks are visible, similar to our
simulation for $\kappa=0.5$ in Fig.\,\ref{fig:RRS38Cor}. For GI GaAs QWs with mixed crystal barriers \cite{OrschelAPL93,JahnJAP95,KopfAPL91} normally more than
2 ML peaks are present. This indicates that for mixed crystal barriers, the interface disorder is partly uncorrelated, as was also observed in structural
investigations \cite{GottwaldtJAP03}. This finding can be understood considering that a Al$_x$Ga$_{1-x}$As mixed crystal growth surface consists essentially
only of Ga atoms that are mobile. The Al distribution, which determines the interface created after overgrowth with GaAs, is therefore essentially unmodified
during the growth interrupt. The resulting first interface is consequently dominated by short range disorder, and the ML islands formed by the growth interrupt
at the second interface give rise to a larger number of ML peaks due to the missing correlated islands on the first interface.

\subsubsection{Influence of flatness and correlation lengths}

\begin{figure}
\includegraphics*[width=\columnwidth]{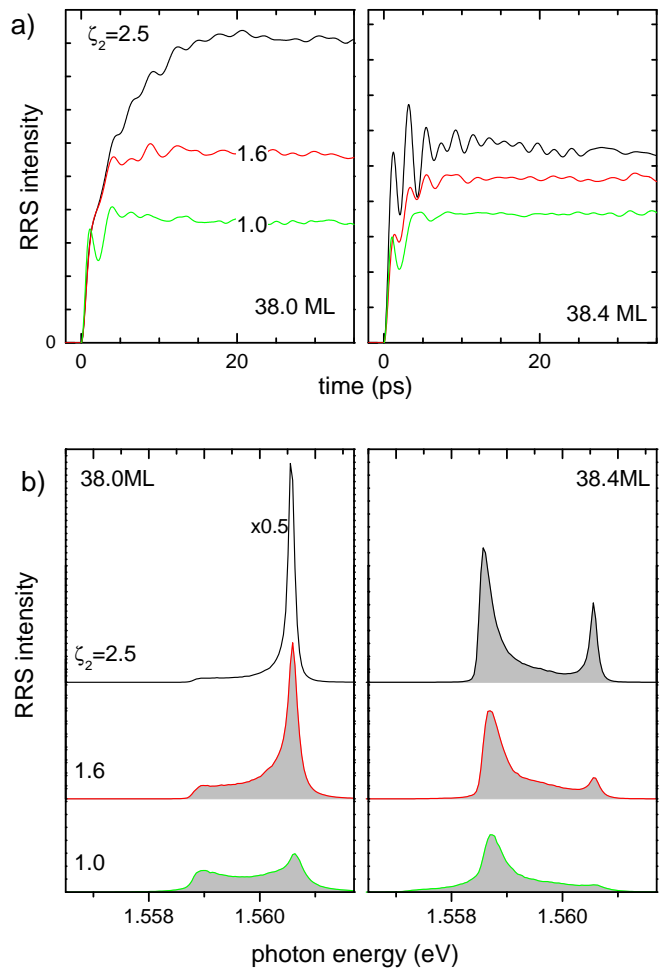}
\caption{\label{fig:RRS38Zdep} Calculated time- (a) or spectrally (b) resolved
RRS intensity versus the flatness $\zeta_2$ for $\zeta_1/\zeta_2=0.75$. Results
for $\Delta=38.0$ (left) and $\Delta=38.4$ (right) are shown. Data are offset
for clarity.}
\end{figure}

\begin{figure}
\includegraphics*[width=\columnwidth]{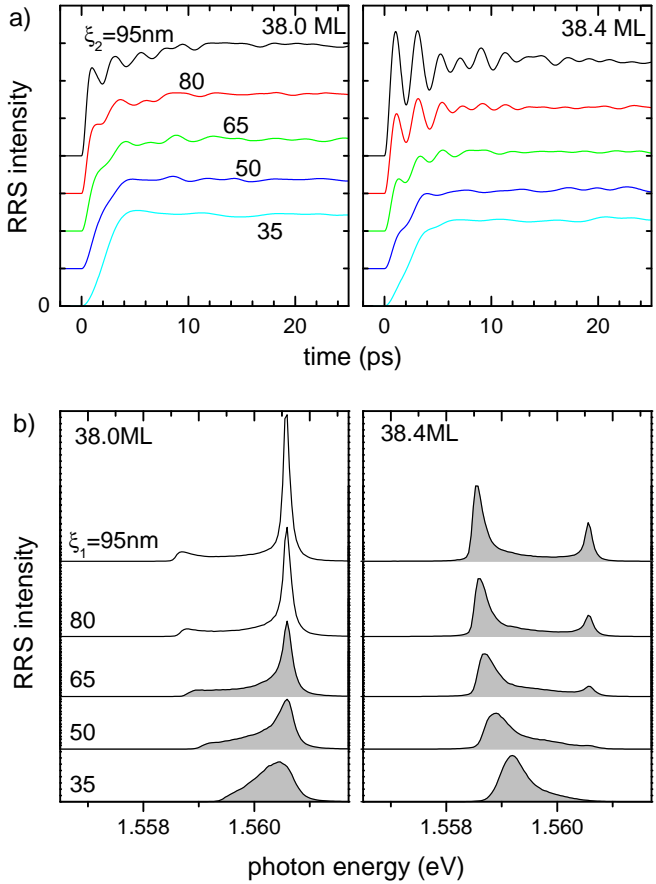}
\caption{\label{fig:RRS38Xdep} Calculated time- (a) or spectrally (b) resolved
RRS intensity versus the correlation length $\xi_1$ for $\xi_1/\xi_2=0.75$.
Results for $\Delta=38.0$ (left) and $\Delta=38.4$ (right) are shown. Data are
vertically offset for clarity.}
\end{figure}

\begin{figure}
\includegraphics*[width=\columnwidth]{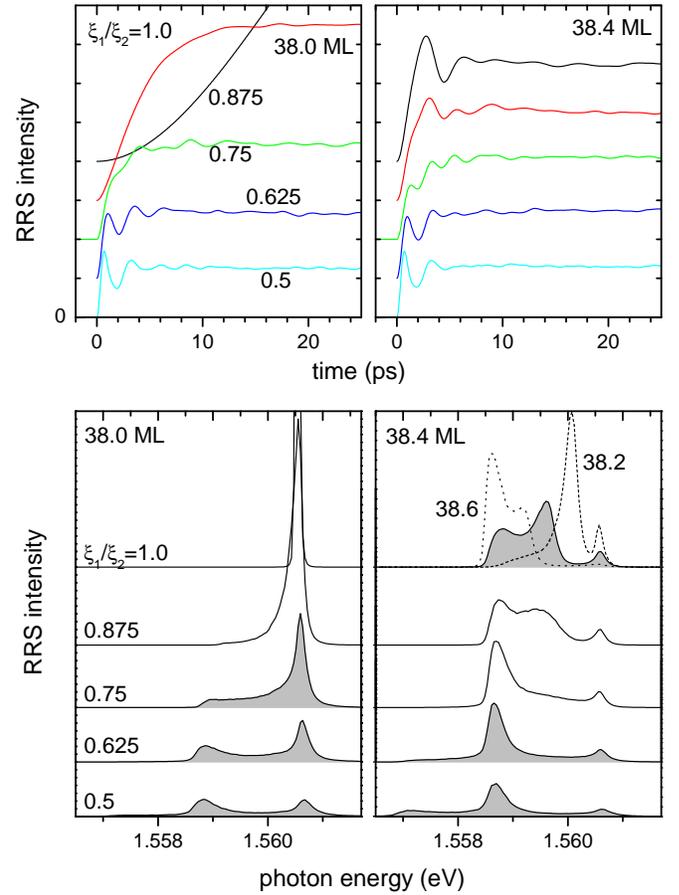}
\caption{\label{fig:RRS38XRdep} Calculated time- (a) or spectrally (b) resolved
RRS intensity versus the correlation length ratio $\xi_1/\xi_2$ for
$\sqrt{\xi_1\xi_2} = 75$\,nm. Results for $\Delta=38.0$ (left) and
$\Delta=38.4$ (right) are shown. Data are vertically offset for clarity. The
dashed and dotted lines in the right lower panel at $\xi_1/\xi_2=1$ are for
$\Delta=38.2$ and $\Delta=38.6$.}
\end{figure}

We now discuss the influence of the flatness $\zeta_1$ and the correlation lengths $\xi_{1,2}$ on the optical response, starting from the parameters in Table
\ref{tab:bestfitpar}(TP). The correlation is taken to be unity ($\kappa=1$), and $\zeta_2/\zeta_1=\xi_2/\xi_1$. We first change interface flatness, from the
value $\zeta_1=1.6$, to a smoother $\zeta_1=2.5$, and rougher ($\zeta_1=1.0$) interface. The calculated RRS is shown in Fig.\,\ref{fig:RRS38Zdep}. Generally,
increasing roughness leads to a spectral broadening of the ML peaks, while their number is hardly affected. In the time-domain, increasing roughness leads to a
decrease of the RRS rise time, and to a suppression of the beating between the ML peaks. From the obtained dependency, we estimate the accuracy of the value in
Table \ref{tab:bestfitpar}(TP) for the flatness $\zeta_1$ to about $\pm30$\%.

Let's now look at the influence of the correlation length $\xi_1$ while keeping the ratio $\xi_1/\xi_2=0.75$ constant. The corresponding simulations are
displayed in Fig.\,\ref{fig:RRS38Xdep}. Increasing this length decreases the lateral quantization energy in the ML islands. At $\xi_1=35$\,nm, the ML peaks are
not developed, since the quantization energy is comparable to the ML splitting. With increasing $\xi_1$, the ML peaks develop into well defined resonances. In
the RRS dynamics, this trend is reflected in the appearance of a well defined temporal beating between the ML peaks. Note the absence of an initial RRS peak
predicted for longer correlation length in a simple disorder potential \cite{SavonaPRB99,KocherscheidtPRB03}. This is related to the perfect interface
correlation (compare also with Fig.\,\ref{fig:RRS38Cor}), effectively canceling the long-wavelength components of the disorder. From the obtained dependency,
we can estimate the accuracy of the value of the correlation length $\xi_1$ in Table \ref{tab:bestfitpar}(TP) to about $\pm15$\%.

Finally, in Fig. \ref{fig:RRS38XRdep} we investigate the influence of the ratio of the
correlation lengths $\xi_1/\xi_2$ for a fixed value of $\xi_1*\xi_2$.  For unity ratio, a difference of
the interface structure exists in our model only for non-integer ML
thickness. For integer $\Delta$ the interface difference is
constant, resulting in a spectrally sharp RRS only broadened by the
short-range disorder $\sigma$. Varying $\Delta$ continuously from
one integer value to the next, rings and stripes of larger thickness form,
gradually getting wider until they entirely cover the plane as
$\Delta$ reaches its next integer value. The rather well defined
ring width leads to a third peak in the spectrum, with an energy
above the 39\,ML peak given by the one-dimensional quantization
energy across the ring. The peak accordingly shifts to lower energy
with increasing $\Delta$.

With increasing deviation of $\xi_1/\xi_2$ from unity, the effect of the interface correlation decreases, so that for $\xi_1/\xi_2=0.5$ more than two ML peaks
are visible, and the RRS dynamics shows an initial peak, similar to the results for decreasing correlation in Fig.\,\ref{fig:RRS38Cor}, and developing into the
case shown in Fig.\,\ref{fig:RRS38Flat}. From the obtained dependency, we can estimate the accuracy of the value in Table \ref{tab:bestfitpar}(TP)for the ratio
$\xi_1/\xi_2$ to about $\pm8$\%.

\subsubsection{Influence of anisotropy and short-range disorder}

The influence of the correlation length anisotropies within
$0.5<\epsilon_\alpha <2$ on the simulation results is not very
pronounced, and is similar to the combined effects of other
parameters. We thus have chosen the isotropic case
$\epsilon_\alpha=1$ in the following. We remark that an anisotropy
is expected due to the surface reconstruction. The effect of such an
anisotropy on the exciton fine-structure was observed
\cite{GammonPRL96}, and an analysis in terms of a correlated
disorder potential in a 5\,nm wide GaAs/Al$_{0.3}$Ga$_{0.7}$As QW
grown without GI revealed\cite{LangbeinPSS00a} a value of
$\epsilon\approx0.6$.

The short range disorder described by $\sigma$ leads to an overall spectral broadening. Its main effect for the parameter set in Table \ref{tab:bestfitpar}(TP)
is to broaden the sharp ML peaks developing for $\Delta$ close to an integer ML thickness. The present value is 0.8\,meV. To interpret this value, we consider
that the volume over which the exciton averages contains about $N=30000$ Ga/Al sites. For a random distribution of $n$ Al sites ($n\ll N$), we would
expect\footnote{As the exciton wavefunction is not constant within the well, a weighted averaging should be performed.} a disorder potential variance of
$\sigma=\sqrt{n}E_c/N$, where $E_c=1.6$\,eV is the band gap difference between GaAs and AlAs at the $\Gamma$ point. From $\sigma=0.8$\,meV we estimate an
average Al concentration of $n/N\approx0.75\%$. This rough estimate is consistent with the expected segregation \cite{JensenJAP99}.

\section{Comparison with the Experiment}

We now compare the calculated optical properties with the measured RRS and
optical density of the sample described as wafer 5 in
Ref.\,\onlinecite{LeossonPRB00}. It contains single GaAs quantum wells,
nominally 39, 30, 23, and 18 ML wide. The sample was grown by MBE at
630$^\circ$C on a two-inch undoped GaAs [100] wafer using 8\,nm AlAs barriers
and 50 nm GaAs spacers between the wells. This barrier design results in a
negligible charging of the QWs because long-lived unpaired charge carriers can
tunnel through the narrow AlAs barriers into the GaAs spacers within
microseconds. Rotation of the substrate was stopped only during the growth of
the wells in order to achieve a continuous variation in well thickness across
the wafer (along the $[110]$ direction) while maintaining a constant barrier
width. Growth rates were calibrated using reflectance high-energy electron
diffraction on a reference wafer. The nominal growth rates were 0.8 ML/s and
0.3 ML/s for GaAs and AlAs, respectively, and a 30\% variation in GaAs growth
rate was observed across the wafer. A standard V/III flux ratio of 8 - 10 was
used. At each position on the wafer the excitonic resonances of the 4 QWs of
different thickness are spectrally well separated, so that by choosing the
excitation wavelength we could efficiently select the excitonic response of a
single QW.

\begin{figure}
\includegraphics*[width=\columnwidth]{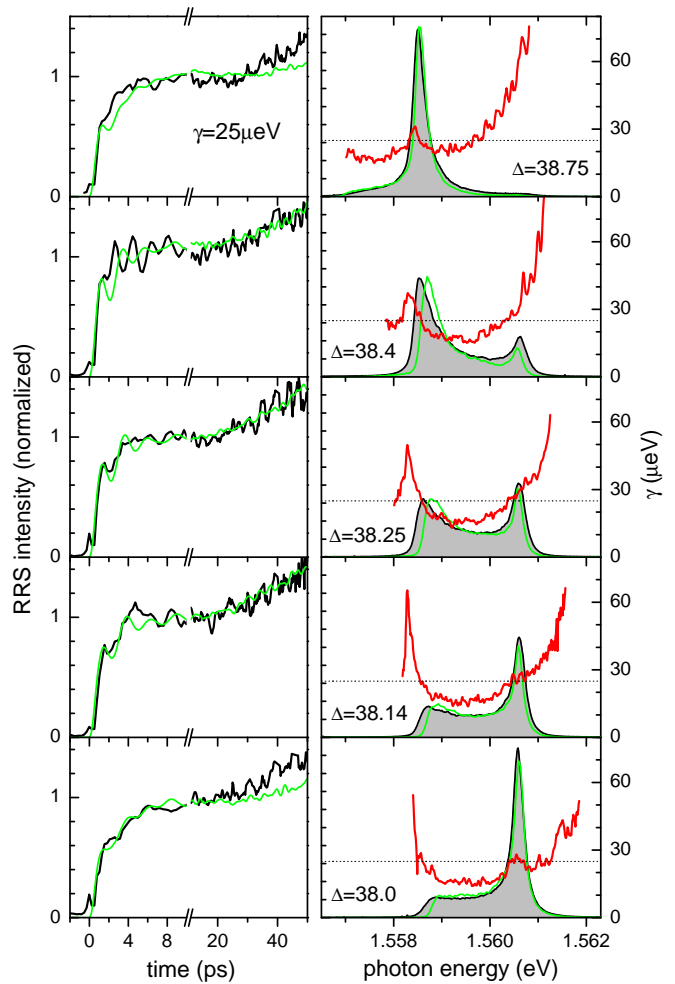}
\caption{\label{fig:ExpRRS} Spectrally and time-resolved RRS intensity for
different QW thicknesses $\Delta$ as indicated. The spectrally resolved RRS is
shown on a constant intensity scale for all $\Delta$, together with the
homogeneous linewidth (HWHM) $\gamma_{\rm ssa}(\omega)$ given in red. The
time-resolved RRS intensity was measured by spectral interferometry with
300\,fs resolution, and is divided through by the exponential decay
$\exp(-2t\gamma)$ using $\gamma=25\,\mu$eV. Note the break in the time axis.
The intensity scale is adjusted for each $\Delta$ to the plateau of the data.
The statistical error due to the finite number of speckles is $\pm$3\%. The
simulated RRS using the ES model (see Section IV.C below) and the parameters of
Table\,\ref{tab:bestfitpar} are superimposed as green lines. The intensities
have been scaled to match the measurements, and the time-resolved data have
been corrected for an average polarization decay rate of $14\,\mu$eV.}
\end{figure}

The spectrally resolved RRS intensity was measured using spectral speckle analysis \cite{KocherscheidtPRB02}, which also determines the spectrally resolved
homogeneous linewidth $\gamma_{\rm ssa}(\omega)$. The samples were kept at temperatures between 1.5\,K and 50\,K. The fundamental hh1-e1 1s exciton resonance
was excited by optical pulses from a mode-locked Ti:Sapphire laser of 0.3\,ps pulse length. The diameter of the excitation spot was 200 $\mu$m. The excited
exciton density was chosen below $10^9\,$cm$^{-2}$ to keep the influence of the density-dependent broadening by exciton-exciton interaction weak. The secondary
emission was passed through an imaging spectrometer and was detected by a nitrogen cooled CCD array with a Voigt spectral response of 20\,$\mu$eV Gaussian and
7\,$\mu$eV Lorentzian full width at half maximum (FWHM). The directional resolution was adjusted to resolve a single speckle, {\it i.e.} to be better than the
diffraction limit of the excited area on the sample.\cite{RungePRB00} All presented data were taken through an analyzer parallel to the linearly polarized
excitation impinging on the sample in Brewster angle, in a directional range centered normal to the sample. The time-resolved RRS was measured using passively
stabilized spectral interferometry \cite{KocherscheidtPSS03,LangbeinSpringer04} -- a simpler approach as compared to actively stabilized versions
\cite{HayesPRB00,BirkedalPRL98}. The time resolution of this measurement is 300 fs, while an uncertainty of $\pm3\%$ in the measured intensity is introduced by
the average over a finite number of speckles. In this case, the diameter of the excitation spot was 100 $\mu$m.

\begin{figure}
\includegraphics*[width=\columnwidth]{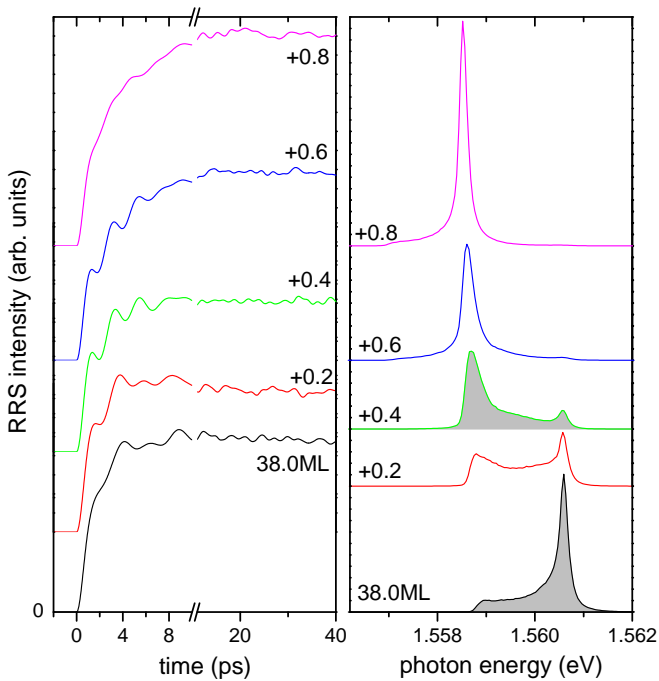}
\caption{\label{fig:RRS38TP} Calculated time- (left) or spectrally (right)
resolved RRS intensity versus $\Delta$ as indicated. Simulation parameters as
in table\,\ref{tab:bestfitpar}. Data are offset for clarity.}
\end{figure}

The measurements were taken at a sample temperature of 1.5\,K. Different positions on the sample along the thickness gradient were investigated. The measured
RRS for varying fractional ML thickness is given in Fig.\,\ref{fig:ExpRRS}. Assuming a linear thickness variation, the observed lateral separation of 3\,mm
between equivalent fractional ML positions of 38 and 39 ML thickness was used to calibrate the thickness gradient. The absolute thickness was fixed by
comparison of the measured RRS spectra with the simulated ones at integer ML thickness (see Fig.\,\ref{fig:RRS38TP}).

The measured RRS shows the formation of ML peaks as described in the previous
sections. The simulated spectra and the dynamics for the parameters of
table\,\ref{tab:bestfitpar} are given in Fig.\,\ref{fig:RRS38TP} and show a
general agreement with the measurements. Some discrepancies in the spectral
shape and in the energy splitting of the ML peaks are still present for
non-integer ML thicknesses. The simulated RRS dynamics reproduces the weak
amplitude of the ML oscillations and their suppression close to integer ML
thicknesses. The simulated oscillation period is about 10 \% longer than in the
measurements, related to the discrepancy in the ML peak splitting. The
deviations could possibly be reduced by further optimizing the model
parameters. However, one qualitative deviation between simulation and
experiment is obvious: the presence of a distribution of polarization decay
rates $\gamma$ within the exciton ensemble. It manifests itself as a non-
exponential decay of the RRS with time, yielding a positive curvature of the
RRS dynamics when corrected for a single exponential decay. This non-
exponential decay was already noted earlier \cite{ KocherscheidtPRB03}, and
attributed to a distribution of $\gamma$ values within the exciton ensemble.
The distribution of $ \gamma$ values is even more directly observable in the
spectral speckle analysis data, where a spectral dependence of the line-width
is found (red curves in Fig. \,\ref{fig:ExpRRS}).

The distribution of decay rates originates from the inhomogeneous nature of the
exciton states. Each localized exciton COM eigenstate is characterized by a
different rate of radiative recombination and phonon scattering to other
exciton states. To model this distribution, one needs first to compute the
exciton eigenstates -- a much more computationally demanding task that could be
avoided in the first part of this work by solving directly the dynamical RRS
equation (\ref{RRS}) versus time, with the drawback of having to use a constant
value of $ \gamma$. Once the exciton eigenstates are known, the distribution of
$\gamma$ can be obtained by computing their radiative recombination and
acoustic phonon scattering rates in first-order perturbation
\cite{KocherscheidtPRB02, MannariniPSS03,ZimmermannBook03}. We have implemented
this approach for the presented disorder model, as we discuss in the next
section. The resulting simulations are superimposed to the measurements in
Fig.\,\ref{fig:ExpRRS} (green curves), and we discuss their comparison with the
measurements after introducing the model in the next section.

\begin{table}[b]
\begin{tabular}{|l|c|c|c|c|} \hline
$\Delta$& $dF/d\Delta$ (meV)& $E_{\rm B}$(meV)&$a_{\rm B}$(nm)&$\sigma$(meV)\\
\hline
38 & 2.1 & 11 & 8 & 0.8\\
30 & 4.5 & 12.4 & 7.5 & 1.7\\
20 & 10.5 & 15 & 6.9 & 4\\
 \hline
\end{tabular}
\caption{\label{tab:mldep} Parameters used in the calculations for different QW
thicknesses.}
\end{table}

\begin{figure}
\includegraphics*[width=\columnwidth]{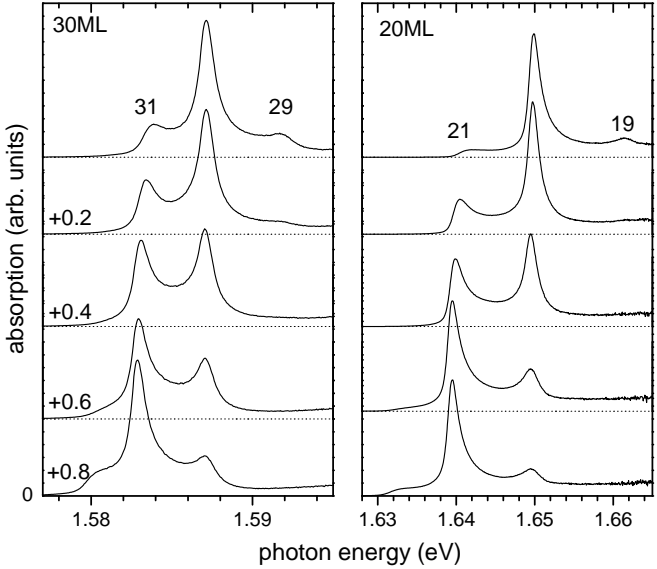}
\caption{\label{fig:PLodW2030} Measured absorption for 20+ and 30+ ML
thickness. Data are deduced from photoluminescence at 50\,K lattice temperature
assuming exciton thermalization to 55\,K.}
\end{figure}

\begin{figure}
\includegraphics*[width=\columnwidth]{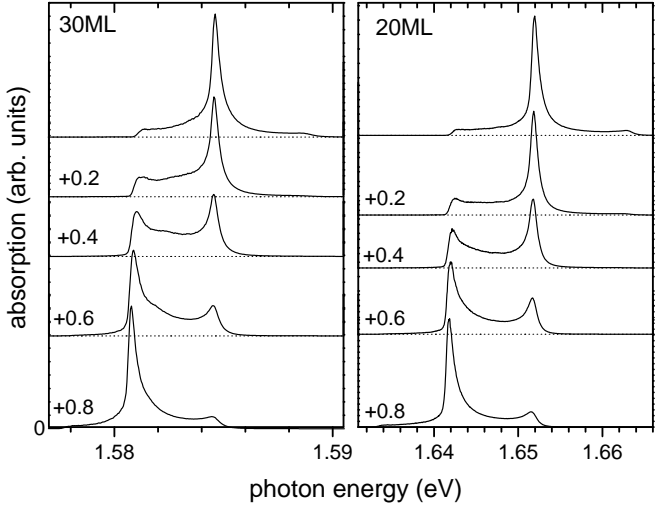}
\caption{\label{fig:REFW2030} Calculated absorption for 20+ and 30+ ML
thickness. Data are offset for clarity.  }
\end{figure}

To evaluate to what extent the present disorder model is applicable to excitons
in QWs of different thicknesses in the same sample -- which we assume to
feature similar interface structures -- we have measured the photoluminescence
(PL) of the QWs of about 30 and 20 ML thickness at a temperature of 50\,K. For
this elevated temperature, we can assume a thermal equilibrium in the exciton
state occupation \cite{GurioliPRB94}, which allows us to determine the spectral
shape of the absorption in a simple way \cite{LeossonPRB00}. The resulting data
(corrected for the temperature induced band-gap shift of 3\,meV) is shown in
Fig.\,\ref{fig:PLodW2030}. In the absorption, up to 3 different ML peaks are
visible. With increasing ML splitting energy, the ML peaks get better defined
due to the smaller localization length $\xi_0$ compared to the constant
interface correlation length. For 20\,ML thickness, the exciton continuum
absorption is merging with the upper ML peak, a clear sign that the rigid
exciton approximation\cite{ZimmermannBook03} will not be of good accuracy. The
exciton binding energies $E_{\rm B}$ for the 30 (20) ML QW are 12 (15) meV,
while the ML splittings are 4 (10) meV, so that already for the 30 ML the
disorder potential amplitude is close to the exciton binding energy. The
exciton binding energies have been determined by a line-shape fit to the
absorption of non-GI reference QWs \cite{LeossonPHD01} and are in agreement
with the values reported in Ref.\,\onlinecite{GurioliPRB93}.

The calculated absorption is given in Fig.\,\ref{fig:REFW2030}. The exciton Bohr radius $a_{\rm B}$ was scaled relative to the 38\,ML simulation according to
$E_{\rm B}^{-1/2}$. All interface parameters are as in table \,\ref{tab:bestfitpar} except $\sigma$, which has been scaled proportional to the ML splitting
energy (see Table\,\ref{tab:mldep}). General agreement between simulation and experiment is found. Quantitative differences are narrower lines and less ML
peaks in the result of the simulation, compared to experimental data. The homogeneous broadening $\gamma$ at 50\,K can be estimated to be below 200\,$\mu$eV,
and should thus not be the main origin of the different line-widths in theory and experiment.

\section{Microscopic treatment of phonon scattering and radiative decay\label{eigenstates}}

We now describe the theoretical approach used to simulate the RRS
signal accounting for a distribution of the polarization decay
rates. We consider two physical mechanisms. First, the radiative
decay of each localized exciton. Second, the polarization decay due
to scattering between different exciton states assisted by acoustic
phonon emission or absorption. The details of how both rates are
modeled for exciton COM eigenstates are described in
Ref.\,\onlinecite{ZimmermannBook03}. The effect of a radiative
renormalization of the states that results e.g. in a polarization
splitting \cite{GammonPRL96,LangbeinPSS00a} is neglected. The
typical splitting energies are similar to the radiative decay rates.
The splitting can lead to an additional slow dynamics of the RRS due to
polarization beating, mainly relevant on a timescale longer than that considered
in the present analysis.

In order to compute the rates it is necessary to know the
eigenstates and energy eigenvalues of the exciton COM motion within
the disorder potential. These are obtained by solving the eigenvalue
problem given by the COM Schr\"odinger equation (\ref{RRS}), which
reads
\be \left(-\frac{\hbar^2}{2M}\nabla^2+V({\bf R})\right)\psi_j({\bf
R})=E_j\psi_j({\bf R})\,, \label{Schr} \ee
where $\psi_j({\bf R})$ are the COM
eigenfunctions and $E_j$ the corresponding energy eigenvalues. Eq. (\ref{Schr})
is solved numerically on the same grid used for simulating the time-dependent
equation (\ref{RRS}). When neglecting decay rates, we can express exactly the
Green's function of Eq. (\ref{RRS}) in terms of its eigenvalue decomposition
\be G({\bf R},{\bf R}^\prime,t)=\sum_j\psi^*_j({\bf R}^\prime)\psi_j({\bf
R})\exp\left(- \frac{iE_j}{\hbar}t\right)\,. \label{green} \ee
From this Green's function, the solution $P({\bf R},t)$ of (\ref{RRS}) for an
arbitrary input field $E_{\rm in}({\bf R},t)$ can be computed. The advantage of
this approach is that we can generalize the solution by introducing decay rates
$\gamma_j$ for each exciton COM eigenstate. The corresponding COM Green's
function can be rewritten as
\be G({\bf R},{\bf
R}^\prime,t)=\sum_j\psi^*_j({\bf R}^\prime)\psi_j({\bf R})\exp\left(-
\frac{iE_j}{\hbar}t-\gamma_jt\right)\,. \label{green2} \ee
This expression can no longer be related to the solution of Eq.
(\ref{RRS}). However, as we will see later in this section, once the
decay rates have been accurately modeled, it gives a better account
of the observed RRS dynamics. As for the time-propagation
simulations we use a delta-pulse excitation field at $\bk_{\rm in}=0$,
$E_{\rm in}({\bf R},t)=E_{\rm in}\delta(t)$, corresponding to an
excitation beam impinging on the sample at normal incidence. For
this configuration, the RRS amplitude resulting from the
simulation can be rewritten as
\be \label{eqn:Pes}
E_{\rm out}(\bk,t)=\sum_j\psi^*_j(\bk)\psi_j(\bk_{\rm in})\exp\left(-
\frac{iE_j}{\hbar}t-\gamma_jt\right)\,, \ee
where
\be \psi_j(\bk)=\frac{1}{\sqrt{A}}\int d{\bf R}\psi_j({\bf
R})\exp(i\bk\cdot{\bf R})\,, \ee
 is the Fourier transform of the
exciton COM eigenfunction, $A$ being the simulation area.

\subsection{Radiative rates}

The radiative polarization decay rates $r_j$ are computed from the
dipole matrix element, according to the Fermi golden rule. Following
Zimmermann {\em et al.}\cite{ZimmermannBook03} they are
\be
r_j=\frac{\Gamma_0}{A}\sum_\bk\frac{\Theta(k_0^2-k^2)}{k_0\sqrt{k_0^2-k^2}}(k_0^2-k^2/2)|\psi_j(\bk)|^2\,.
\label{radiative}\ee
Here $A$ is the area of the simulation domain, $\bk$ is as usual the in-plane wavevector, whereas $k_0=\sqrt{\epsilon_b}E_x/(\hbar c)$ is the wavevector
corresponding to the {\em light cone}, defining the boundary of the in-plane momentum region where an exciton state can decay into a photon mode radiating in
the $z$-direction. Because of energy and momentum conservation, excitons at $k>k_0$ result into a surface polariton mode characterized by an evanescent light
wave in the vertical direction. The quantity $\Gamma_0$ is the radiative rate of a plane-wave exciton state having zero COM momentum. Expression
(\ref{radiative}) has been averaged over the azimuthal angle and over the two possible polarization directions of the emitted electromagnetic field. This
average is a good approximation, despite the irregularly shaped COM eigenfunctions, as the eigenstates relevant to the optical response are localized on a
spatial range much shorter than the optical wavelength in the medium, which for GaAs is in the range of 250 nm. Consequently, the wavefunctions are
approximately isotropic for $k<k_0$. For the numerical calculations, we have used $\Gamma_0=34~\mu\mbox{eV}$, which is estimated\cite{AndreaniReview95} for a
11 nm GaAs/AlAs QW heavy-hole exciton having an exciton binding energy of 11\,meV.

\subsection{Phonon rates}

The phonon scattering rates are computed accounting for the exciton
deformation potential interaction with longitudinal acoustic
phonons. A detailed account of this calculation is again found in
the Review by Zimmermann {\em et al.}\cite{ZimmermannBook03} and we
report here just the essential expressions. Fermi golden rule gives
the following expression for the polarization scattering from state
$j$ to state $l$ by emission or absorption of a phonon
\begin{eqnarray}
p_{lj}=\frac{2\pi}{\hbar}\sum_{\bf q} |t_{lj}^{\bf q}|^2
\big[(n_q+1)\delta(E_l-E_j+\hbar\omega_q)\\ \nonumber  +
n_q\delta(E_l-E_j-\hbar\omega_q)\big]\,, \label{phFGR}
\end{eqnarray}
where ${\bf q}$ is the phonon momentum in three dimensional reciprocal space,
$n_q$ is the Bose thermal distribution of phonons, $\omega_q=v_sq$ is the
phonon dispersion, $v_s$ being the sound velocity. The quantity $t_{lj}^{\bf
q}$ is the scattering matrix element of the deformation potential Hamiltonian
which, after introducing the factorization ansatz between relative and COM
exciton motion, is given by
\begin{eqnarray}
t_{lj}^{\bf q} = \sqrt{\frac{\hbar\omega_q}{2v_s^2\rho_mV}}(\psi_j\psi_l)_{{\bf
q}_{||}}\big[ D_eK_e(q_z)\chi(q_{||}/\eta_e)\\\nonumber
+D_hK_h(q_z)\chi(q_{||}/\eta_h)\big]\,.
\end{eqnarray}
Here, $\rho_m$ is the mass density of the material, $D_{\alpha}$ are the deformation potentials for electron and hole, $q_z$ and ${\bf q}_{||}$ are
respectively the $z$ and in-plane components of the phonon momentum, and $\eta_\alpha=(m_e+m_h)/m_\alpha$ are mass factors. The functions $K_\alpha$ and $\chi$
are overlap integrals between the phonon wavefunction and, respectively, the confinement functions in the $z$ direction and the relative 1$s$ exciton
wavefunction. The expression \be (\psi_j\psi_l)_{{\bf q}_{||}}=\int d{\bf R}\psi_j^*({\bf R})\exp(-i{\bf q}_{||}\cdot{\bf R})\psi_l({\bf R})\,, \ee is the
corresponding overlap integral for the in-plane COM wavefunctions. After summing Eq.\,(\ref{phFGR}) over the $q_z$ variable we find for the scattering rates
\be p_{lj}=\frac{2L_z}{\hbar^3v_s^2}|\Delta E_{jl}|n(\Delta E_{jl})\sum_{{\bf
q}_{||}}\frac{|t_{lj}^{{\bf q}_{||},\bar{q}_z}|^2}{\bar{q}_z}\,, \ee
where
$\Delta E_{jl}=E_l-E_j$, $\bar{q}_z=\sqrt{(\Delta E_{jl}/\hbar
v_s)^2-q_{||}^2}$ is the restriction imposed by energy-momentum conservation,
$L_z$ is the size of the integration domain in the $z$ direction (which is
canceled when plugging in the explicit expression for the scattering matrix
element), and $n(E)=(\exp(E/(k_BT)-1)^{-1}$ is the Bose energy distribution.

The overall decay rate for a given exciton state $j$ is then given by the sum
of its radiative decay rate and all out-scattering phonon rates $p_{lj}$
\be \gamma_j=r_j+p_j, \qquad p_j=\sum_l p_{lj}\,, \ee
where we have denoted the overall phonon outscattering rate of the $j$-th state
by $p_j$. For the simulations, we use the deformation potentials values $D_e=7$
eV, $D_h=3.5$ eV, the mass density value $\rho_m=5.37~\mbox{g}~\mbox{cm}^{-3}$,
and the sound velocity $v_s=5.33\times10^5~\mbox{cm}~\mbox{s}^{-1}$, typical
for GaAs bulk material.\cite{SiantidisPRB01} The electron and hole effective
masses were respectively $m_e=0.07$ and $m_h=0.18$ times the bare electron
mass.

\begin{figure}
\includegraphics*[width=\columnwidth]{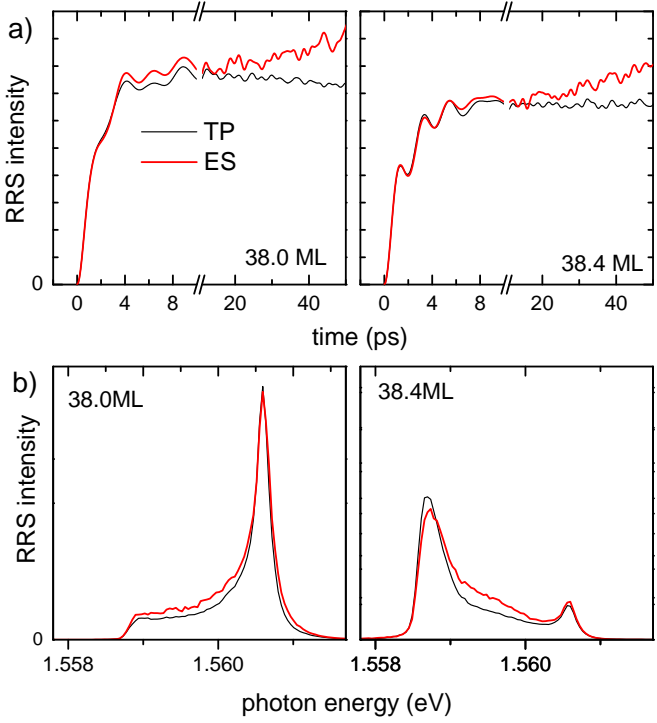}
\caption{\label{fig:ESComp} Calculated time- (a) or spectrally (b) resolved RRS
intensity using the time propagation (black) and the eigenstate calculation
(red). The eigenstate dynamics has been corrected for an average polarization
decay rate $\gamma=14\,\mu$eV. Note the break in the time axis. The spectrally
resolved data are normalized to the spectrally integrated intensity of the
38.0\,ML data. Disorder parameters are given in
table\,\ref{tab:bestfitpar}:TP.}
\end{figure}

\subsection{Simulations: Ensemble averages}

We can now compare the results of the ES calculation (Eq.\,(\ref{eqn:Pes})) with the one of the TP calculation (Eq.\,\ref{RRS}). Using the time-propagation
disorder potential parameters given in Table\,\ref{tab:bestfitpar}, we show in Fig.\,\ref{fig:ESComp} the time-resolved and spectrally resolved RRS intensity
for $\Delta=38.0$ and 38.4\,ML calculated with both methods. The time-resolved RRS calculated in the TP and ES model are nearly identical at early times
$t<10$\,ps. This is because all relevant decay rates $\gamma_j$ are much smaller than the inhomogeneous broadening, so that the nonuniform polarization decay
among the eigenstates is irrelevant ($\exp(-t(\gamma_j-\gamma))\approx1$), and the RRS dynamics is determined by the energy-level distribution associated with
the disordered exciton eigen-states \cite{KocherscheidtPRB03}. For later times, the energy-level distribution creates a constant RRS signal, given by the
incoherent sum of the intensities emitted by the individual exciton states, as can be seen in the TP results. In this regime the decay rate distribution is
relevant since $\exp(-t(\gamma_j-\gamma))$ deviates significantly from unity, yielding a non-exponential RRS decay. The deviation from a single exponential
decay in the ES calculation is similar to the one measured in the time-resolved RRS (Fig.\,\ref{fig:ExpRRS}), indicating that the ES model is describing its
main physical origin.

\begin{figure}
\includegraphics*[width=\columnwidth]{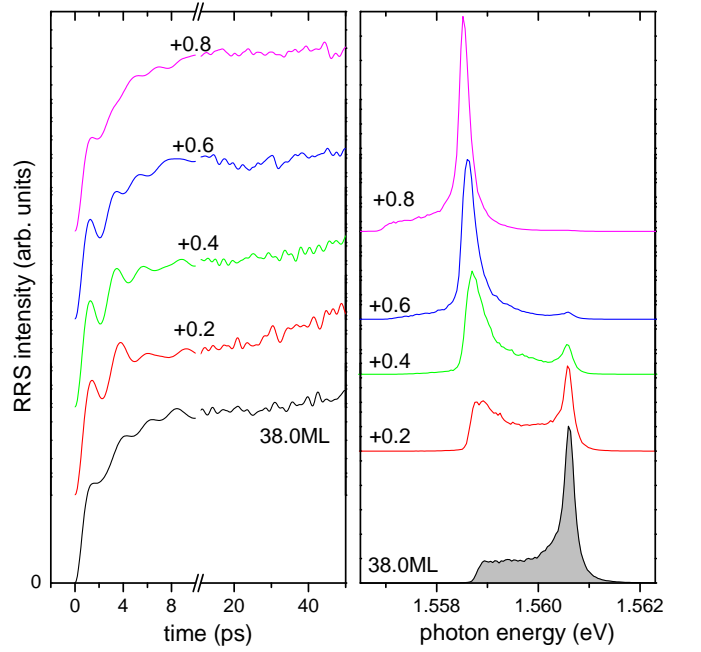}
\caption{\label{fig:RRSESfit} Calculated time- (left) or spectrally (right)
resolved RRS intensity versus $\Delta$ as indicated. The RRS dynamics has been
corrected for an average polarization decay rate $\gamma=14\,\mu$eV. Eigenstate
simulation with parameters as in Table\,\ref{tab:bestfitpar}. Data are offset
for clarity.}
\end{figure}

Comparing the spectra of the TP and the ES calculation, some
difference are apparent, especially for energies between the ML
peaks - the ES calculation tends to result in more spectral
broadening. This difference relates to the distribution of phonon
and photon related dephasing rates. Exciton states energetically in
between the ML peaks are likely to be excited states of a ML trench,
which have rather small radiative rates. Since their phonon
scattering rates are still smaller than the average radiative rates,
their dephasing rate is smaller than average, resulting in a longer
and thus spectrally stronger RRS emission as compared to the TP
result. To reproduce the experimentally measured RRS spectra with
the ES calculation, we needed therefore to adjust the disorder model
parameters slightly, as detailed in the ES row of
table\,\ref{tab:bestfitpar}. The resulting calculated RRS spectra
and dynamics are given in Fig.\,\ref{fig:RRSESfit} and show a
general agreement with the measurements. However, the calculated
average polarization decay rate of 14\,$\mu$eV is lower than the
measured one of 25\,$\mu$eV. Such a deviation was noticed already in
previous works \cite{KocherscheidtPRB02}, and can have different
origins. On the theoretical side, the deformation potentials, which
enter the phonon scattering rates quadratically, are only known up
to a factor of two. The radiative rate $\Gamma_0$ can be influenced
by the dielectric surrounding including dipole-dipole interaction
between the exciton states, and also directly by the disorder
potential if the rigid exciton approximation breaks down
\cite{ZimmermannBook03,GrocholPRB05} resulting in a modified
electron-hole overlap. On the experimental side, there could be an
influence of residual exciton-exciton scattering, the influence of
spectral diffusion \cite{TurckPRB00}, and also a systematic error
due to the finite spectral resolution. Measurements of the dephasing
time of exciton ensembles in similar structures by four-wave mixing
\cite{ErlandPRB99} showed a polarization decay comparable to the one
measured here. We have made no attempt to modify the model
parameters towards a better agreement with the experiment since we
are mainly concerned with the effect of the disorder potential in
this work.

\begin{figure}
\includegraphics*[width=\columnwidth]{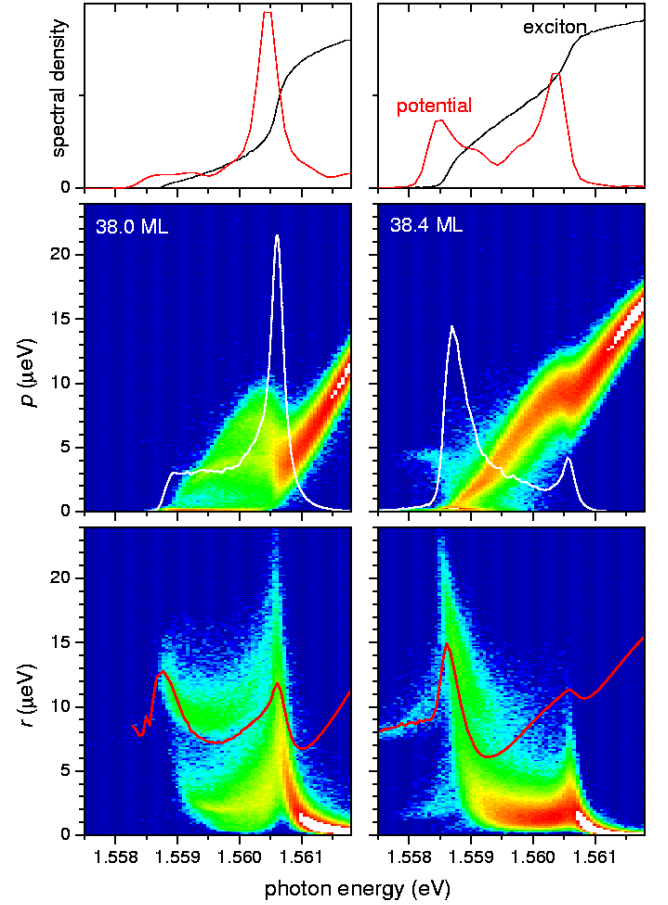}
\caption{\label{fig:ESgdis} Probability distribution of phonon
scattering rates $D_p(p,\omega)$ and radiative decay rates
$D_r(r,\omega)$ for exciton states in a QW of 38.0 and 38.4\,ML
thickness. Other parameters as in Table\,\ref{tab:bestfitpar}.
Logarithmic color scale over three decades. The RRS spectra are
superimposed as white lines, and the average decay rate
$\gamma(\omega)$ as red lines. The top plots show the exciton
density of states {\bf $D_{\rm X}(\omega)$} and the potential
density {\bf $D_V(\omega)$}.}
\end{figure}

Having at hand the exciton eigenstates with their radiative emission
and phonon-assisted relaxation rates, we can investigate in detail
their properties at different energies in the disorder potential. We
do this in the following for $\Delta=38.0$ and 38.4\,ML. The
calculated spectral distributions of the density of states (DOS) and
the phonon and radiative rates for low temperature (T=1.5\,K) are
given in Fig.\,\ref{fig:ESgdis}. The exciton DOS {\bf $D_{\rm
X}(\omega)$} is compared in the top panel to the normalized spatial
potential density {\bf $D_V(\omega)$}. For long potential
correlation lengths, the lateral quantization energies are
negligible and {\bf $D_{\rm X}(\omega)$} should be just given by the
convolution of {\bf $D_V(\omega)$} with the 2D exciton DOS
$\theta(\hbar\omega-E_0)\frac{M}{\pi\hbar^2}$:
\be D_{\rm
X}(\omega)\approx\frac{M}{\pi\hbar}\int_{\omega'=-\infty}^{\omega}D_V(\omega')d\omega'\ee
This general behavior is visible also in the calculated data, but
due to the finite lateral quantization energies, an additional
blueshift of {\bf $D_{\rm X}(\omega)$} is observed.

The distributions of the scattering rates are more interesting, as they provide information about the character of the states. At a temperature of 1.5\,K, the
thermal energy $\kB T \approx 0.1\,$meV is much smaller that the ML splitting energy of about 2\,meV, so that phonon absorption processes are not significant.
In this case, the phonon-assisted transitions are directed towards lower-lying exciton states with significant spatial overlap. In a ML island, the local
exciton ground state has thus a negligible $p$, while the first excited state acquires some $p$, and the second excited state even more. The radiative coupling
shows the opposite behaviour: The ground state has a nodeless wavefunction and thus a large $k<k_0$ component, resulting in a large $r$ (see
Eq.\,(\ref{radiative})). Excited states have additional nodes in the wavefunction, and thus suppressed $k<k_0$ components. The calculated distributions
$D_p(p,\omega)$ and $D_r(r,\omega)$ of $p$ and $r$, respectively, are shown in the lower four panels of Fig.\,\ref{fig:ESgdis}. For $\Delta=38.0$, the 39 ML
regions have only a small spatial coverage, and thus form sub-wavelength sized trenches surrounded by 38\,ML regions. Most exciton states in the spectral
region below the 38\,ML potential energy $F(38)$ are thus localized in individual trenches of 39\,ML thickness and show the discussed behaviour. The states of
lowest energy ($\omega<1.559$\,eV) are dominantly ground states, showing a large $r$ and vanishing $p$. In the region 1.559\,eV$<\omega<$1.560\,eV, the first
two excited states (p-like) show up, with a 3-5 times reduced $r$, but a significant $p\approx 3\,\mu$eV. For 1.5596\,eV$<\omega<$1.5605\,eV, also higher
excited states are present as indicated by a feature of $D_p(p,\omega)$ at $p\approx 7$\,$\mu$eV, and a broadening of $D_r(r,\omega)$ of the excited states.
The ground states show a non-monotonic dependence of $r$ on the transition energy: decreasing with increasing energy up to halfway to the energy of the 38\,ML
barrier, and then increasing again. This can be understood considering the corresponding evolution of the wavefunction extension: The state energy separation
to the 39\,ML energy (1.5585\,eV) can be interpreted as quantization energy $E_Q$ of the state in the trench. For infinite barriers and a square trench of size
$L$, we find $E_{\rm Q}=(\pi\hbar/L)^2/M$, yielding 0.3\,meV for $L$=100\,nm. Small quantization energies thus correspond to large trench sizes with large
wavefunctions having large $k<k_0$ components. With decreasing trench size, the wavefunction decreases in size, until $E_{\rm Q}$ reaches about half the
barrier height. With further decrease in trench size, the increasing penetration of the wavefunction into the lateral barrier leads to an increase of the
wavefunction size and thus of the $k<k_0$ component. This behaviour was previously predicted \cite{SugawaraPRB95,AndreaniPRB99} and experimentally
verified\cite{HoursPRB05}.

The energy-dependent averaged linewidth
\be
\gamma(\omega)=\langle\gamma_j\rangle_{E_j=\hbar\omega}
\label{gammaomega}
\ee
is given in the lower panel of Fig.\,\ref{fig:ESgdis}. It shows a dependence similar to
the one measured by spectral speckle analysis (see Fig.\,\ref{fig:ExpRRS}), with
maxima close to the integer ML energies due to the large radiative rates of the
corresponding weakly confined exciton states. In this comparison, we have to keep in
mind that the weighting of the states in the average linewidth determined by spectral
speckle analysis \cite{KocherscheidtPRB02,LangbeinPSS02c,LangbeinHabil02} is not
just uniform, as assumed in Eq. (\ref{gammaomega}), but each state is weighted by
its RRS intensity, scaling like $r^2/(r + p)$. This weighting leads to a measured
dephasing which is typically by some 10\% larger that the unweighted average $
\gamma(\omega)$. Yet, this weighting cannot explain the observed factor of two
between experiment and prediction.

For a direct quantitative comparison of the simulated RRS spectra and dynamics with
the measurements, we show in Fig.\,\ref{fig:ExpRRS} the ES simulation results for the
exact ML thicknesses of the measurements superimposed on the measured data. The comparison shows good overall quantitative agreement, in the general shape of both spectral and time-resolved data and, particularly, in the amplitude of the beatings in the time-resolved data at non-integer ML thickness. We find the following deviations
significantly above the statistical noise, which we believe are not related to a lack of
fine-tuning of the model parameters:

i) The oscillation beat frequency and the splitting of the monolayer energy are about 10
\% smaller in the simulation, even though the energy shift due to a thickness change
of one ML is perfectly reproduced.  In the model, the smaller splitting for non-integer
ML thickness is a natural consequence of the lateral quantization in the islands of finite
size. It is also instructive to observe that this effect is less pronounced in the
simulation of the thinner QWs (see Fig.\,\ref{fig:REFW2030}) since the in-plane size of
the ML islands is constant, while the ML splitting increases as the QW thickness decreases. In the corresponding
experiment instead (Fig.\,\ref{fig:PLodW2030}), the effect is more pronounced. One
possible origin of this deviation could be the rigid exciton approximation. Removing this
approximation, exciton states localized in the small monolayer islands would display a
larger electron-hole coulomb interaction\cite{GrocholPRB05}, leading to a shift of the
lower ML peak to lower energy. Unfortunately it is numerically prohibitive to perform
the necessary ensemble averages without this approximation. The observed deviation
of about 0.3\,meV is a small correction relative to the exciton binding energy of 11\,
meV, and lies within the expected magnitude of this effect. Another possible source of discrepancy are finer
details of the heterointerface structure that are not accounted for in our disorder model. For
example a non-Gaussian spatial correlation could give rise to a larger spread of island
sizes, which reduces the lateral quantization in the largest islands. Extending the model in such direction is a possible matter of future development, but we feel it should be supported by
more extensively sampled experimental data.

ii) The long-time nonexponential RRS dynamics is not well reproduced close to integer
ML thickness. Generally, this dynamics is given by the distribution of dephasing rates
across the state ensemble, and also, since we excite and detect light linearly polarized
along $[1\bar{1}0]$, by the distribution of the fine-structure splitting and its
orientation\cite{LangbeinPSS00a}. We have discussed the dephasing rate distribution
already in the previous paragraph. The fine-structure splitting would lead to an initial
decay of the signal within a time given by the inverse average splitting energy, typically
in the 20-50 ps regime\cite{LangbeinPSS00a}. This would result in a stronger non-
exponential RRS decay, similar to the deviation observed. The reduced importance of
this effect for non-integer ML thickness could be due to the alignment of the
fine-structure along the $[1\bar{1}0]$ direction in the (anisotropic) ML islands. A
numerical calculation of the fine-structure splitting in the present model is only feasible
in a diagonal approximation in the eigenstates, due to computational limitations. This
extension of the model, together with a polarization direction dependent RRS dynamics
measurement, could be an interesting future investigation.

\begin{figure}
\includegraphics*[width=\columnwidth]{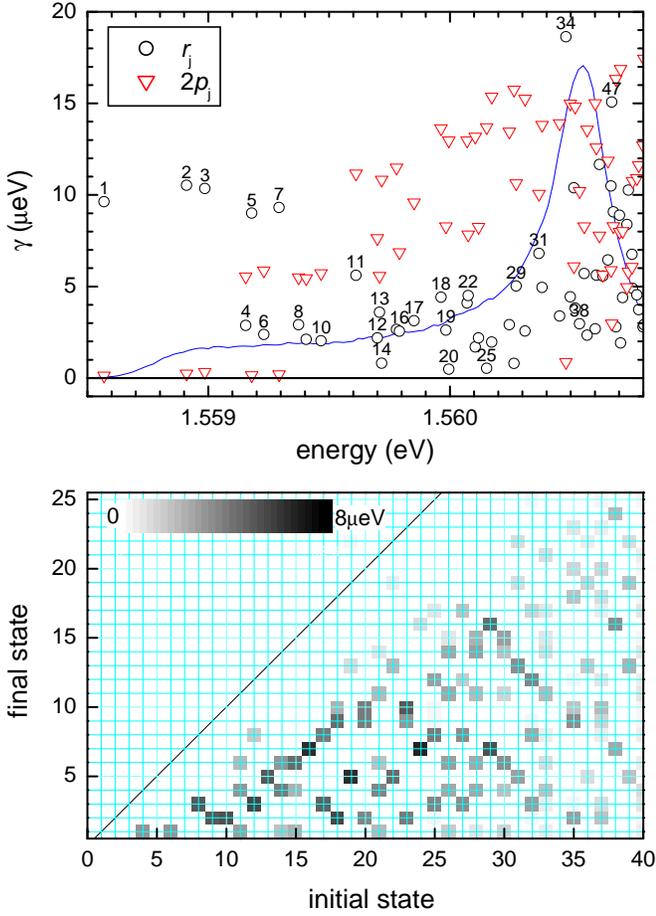}
\caption{\label{fig:C5Enstates} Excitonic states in a single
disorder potential realization $V_{\rm I}(\bR)$ of (500\,nm)$^2$
with periodic boundary conditions, using the parameters of
table\,\ref{tab:bestfitpar} and $\Delta=38.0$. Top: Phonon
scattering ($p_j$) and radiative decay rates ($r_j$) of the exciton
states $j$ with energy $E_j$. The numbered states $j=1,2,...$ have
ascending energy ($E_j<E_{j+1}$). The ensemble averaged RRS spectrum
is given (blue line) for comparison. Bottom: Phonon scattering rates
$p_{lj}$ from the state $j$ to the state $l$ at T=1.5\,K. Linear
grey scale as displayed.}
\end{figure}

\begin{figure}
\includegraphics*[width=\columnwidth]{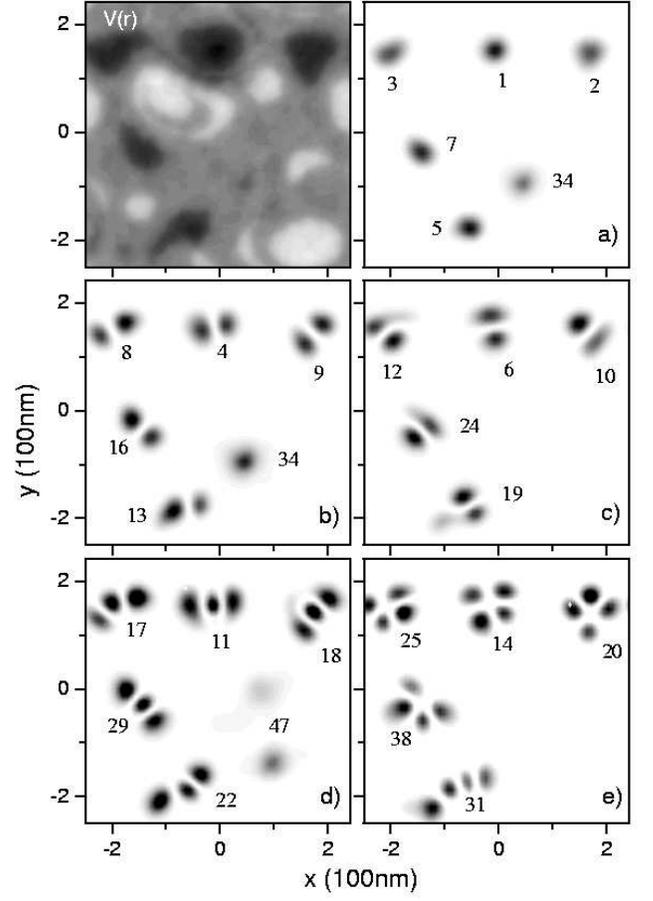}
\caption{\label{fig:C5wavefunc} Disorder potential $V_{\rm I}(\bR)$ (top left)
on a scale from 1.5575 (white) to 1.5635\,eV (black). a)-e) Probability
distributions $|\Psi_\alpha|^2$ of selected excitonic states as labeled. The
grey scale is from zero (white) to a) 800, b,c) 500 and d,e) $300\,\mu$m$^{-2}$
(black).}
\end{figure}

\begin{figure}
\includegraphics*[width=\columnwidth]{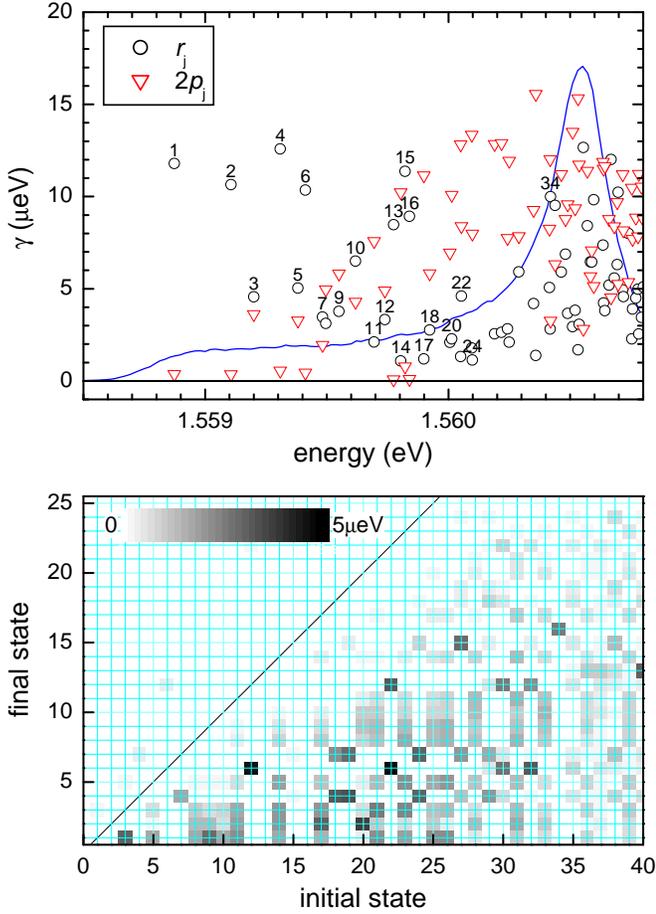}
\caption{\label{fig:C4Enstates} As Fig.\,\ref{fig:C5Enstates} for a different
disorder potential realization $V_{\rm II}(\bR)$. }
\end{figure}

\begin{figure}
\includegraphics*[width=\columnwidth]{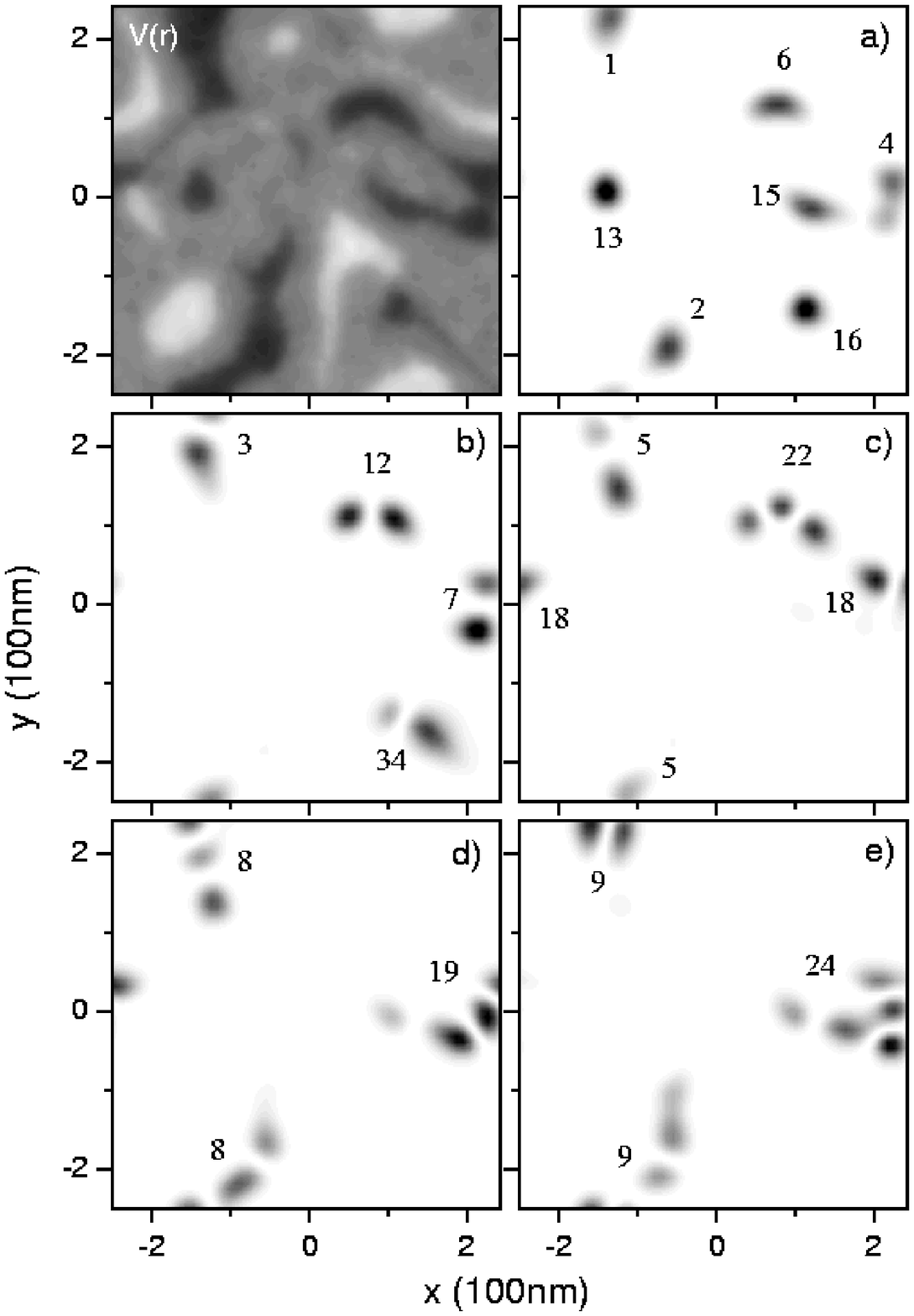}
\caption{\label{fig:C4wavefunc} As Fig.\,\ref{fig:C4wavefunc} for the disorder
potential realization $V_{\rm II}(\bR)$. }
\end{figure}

\subsection{Simulations: Single realizations}

Having described in the previous section ensemble averaged properties, we turn
now to individual excitonic states in specific disorder realizations. These
results are important for comparison with the spectroscopy of individual
excitonic states performed using high spatial resolution \cite{WuPRL99,
GuestScience01,LangbeinPRL05}. The simulation area used in our calculations of
(500\,nm)$^2$ is comparable to the typical sizes of the
regions probed in this kind of experiments, so that the measured and simulated exciton state ensembles are
similar.

We have chosen the thickness $\Delta=38.0$ for the simulations as for this fractional thickness well defined trenches of 39ML thickness are formed in the
38\,ML plateau. Two realizations I \& II of the disorder are discussed in the following, having trenches of more circular or more elongated and irregular
shape, respectively. The simulated potential landscapes $V_{\rm I,II}(\bR)$ are shown in the top left panel of Fig.\,\ref{fig:C5wavefunc} and
Fig.\,\ref{fig:C4wavefunc}. The remaining panels of Fig.\,\ref{fig:C5wavefunc} and Fig.\,\ref{fig:C4wavefunc} display the square modulus of some exciton COM
wavefunctions selected among the lowest energy eigenstates. Each of these states is, in general, well localized in a local minimum of the disorder potential.
We have therefore chosen to plot in each panel several states having similar features, according to the discussion below. In the plots, these excitonic states
are labeled by an eigen-state number $j$, sorted in order of ascending eigen-energy $E_j$. Each state is characterized by its radiative and phonon scattering
rate $r_j$ and $p_j$, which are plotted in Fig.\,\ref{fig:C5Enstates} and Fig.\,\ref{fig:C4Enstates}, respectively. From the mutual phonon scattering rates
$p_{lj}$ we can identify corresponding ground and excited states of individual trenches.

We start the discussion with realization I. The state 1 is the global exciton
ground state of the system and thus also a local ground state. It shows a large
$r\approx 10\,\mu$eV (corresponding to a dipole moment $\mu\approx 105\,$Debye
as given by $\mu=\sqrt{6\pi\hbar r \epsilon_0 c^3 n^{-1} \omega^{-3}}$ using
the refractive index $n\approx3.5$), and negligible phonon scattering $p_1$, as
expected. States 2,3,5,7 show a similar behaviour, and should thus be
attributed to local ground states. States 4,6 instead have a reduced $r\approx
3\,\mu$eV and a significant $p\approx 3\,\mu$eV. They are excited states of
state 1, as can be seen from the phonon scattering matrix $p_{lj}$ given in the
lower panel of Fig.\,\ref{fig:C5Enstates}, which shows sizable $p_{14}$ and
$p_{16}$, but negligible $p_{24}$, $p_{34}$, $p_{26}$, $p_{36}$. From the
$p_{lj}$, we find also that the states 11,14,15 are further excited states of
state 1, showing also relaxation into the states 4,6. Similarly, states
9,10,18,20,23 are excited states of 2, and 8,12,17,25,27 are excited states of
3. Energetically close to the 38\,ML potential the local ground state 34 is
formed, which has an exceptionally large $r\approx 19\,\mu$eV, indicating its
large spatial extension.

This discussion can be verified by observing the spatial wavefunctions of the
various states, which are displayed in Fig.\,\ref{fig:C5wavefunc}. The states
1-4-6 represent a s-p$_x$-p$_y$ sequence, where the s,p,d,e,.. states refer to
the sequence of quantized state groups in a two-dimensional harmonic oscillator
which have the degeneracies 1,2,3,4,.. . The higher excited states 11,14 (and
15, not shown) are d-states with 2 nodes, but are already quite deformed due to
the non-perfect circular symmetry of the potential trench. Similar sequences
are observed for the states 2-9-10-18-20-23, 3-8-12-17-25-27 and 7-16-24-29-38.
In the sequence 5-13-19-22-31 the asymmetry of the trench is so strong that the
d and e states are strongly mixed. The state 34 is a rather large state
localized by a 39\,ML trench about the size of $a_{\rm B}$, for which the
confinement potential is therefore already reduced due to the averaging over
the exciton relative wavefunction $|\phi({\bf R}')|^2$.

Generally we can say that for realization I, most of the states below the 38\,ML peak are well described by ground and excited states of single rather round ML
island trenches. In realization II, most of the trenches are instead elongated and bending, some even show bifurcations (see Fig.\,\ref{fig:C4wavefunc}).
Accordingly, the states localized in the trenches do not resemble the s-p-d state sequence. Looking at $r$ and $p$ in Fig.\,\ref{fig:C4Enstates}, the states
1,2,4,6,13,15,16 are local ground states. State 1 has the excited states 3,5, and shares the excited states 8,9,10,11 with state 2. This is due to the
connected nature of the trenches, which support multiple local ground states within a trench structure. Such a structure results in a exciton state system with
two metastable local ground states (here 1,2), and excited states delocalized over both ground states (here 8,9,10,11) which allow to coherently couple the two
local ground states by coulomb and exchange interaction with the excited state. Such coherent coupling has been observed in recent literature using nonlinear
spectroscopy \cite{BattehPRB05,UnoldPRL05}.

\begin{figure}
\includegraphics*[width=\columnwidth]{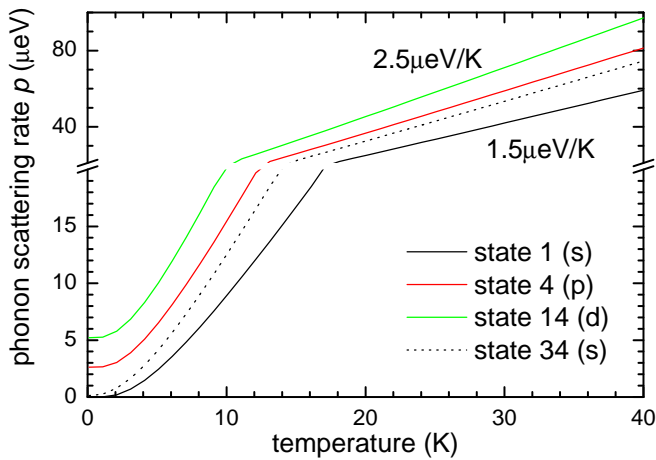}
\caption{\label{fig:ESTdep} Temperature dependence of the phonon
scattering rate $p_j$ for selected exciton states of the disorder
potential realization $V_{\rm I}(\bR)$.}
\end{figure}

The different localization of the states results not only in
different zero temperature dephasing rates, but also in a different
temperature dependence. Strongly localized states show an activated
temperature dependence. Especially the local ground states (s-like)
do show vanishing phonon assisted dephasing at zero temperatures,
and an activated behavior for $T>0$ proportional to the phonon
occupation at the energy of the transition to the first activated
states (p-like). This is exemplified in Fig.\,\ref{fig:ESTdep} for
the states 1(s),4(p),14(d) and 34(s) of the first disorder
realization (see Fig.\,\ref{fig:C5wavefunc}). State 1 shows the
largest activation energy, given by the distance to its first
excited states (4,6) of about 0.6\,meV. The excited states also show
an activation on top of the zero-temperature offset, but with a
smaller activation energy as the higher excited states are typically
energetically due to the finite barrier height. The weakly localized
state 34 shows a weaker activated behavior, similar to what would be
expected for an ideal quantum well, where phonon energies between
about 0.1-1\,meV (for a 11\,nm GaAs QW) can scatter the optically
active $k=0$ state to exciton states of higher $k$
\cite{ThraenhardtPRB00,PiermarocchiPRB96}. These predictions are in
agreement with measurements of the temperature-dependent line-widths
of single exciton states observed in PL experiments
\cite{GammonScience96}. It is interesting to note that not only the
activation energy, but also the linear slope for higher temperatures
($kT>1\,$meV) varies between the states. The localization has thus
an effect on the phonon assisted dephasing even at thermal energies
much larger than the localization energies, creating a significant
distribution in the exciton dephasing rates at elevated
temperatures.

\section{Conclusions}
In conclusion we have presented a model for the disorder of the heterointerfaces in quantum wells that takes into account several important aspects of the
interface formation, like surface diffusion, monolayer island formation, segregation \& interdiffusion, and interface correlation. While not attempting to
model the growth process microscopically, the main features observed in structural investigations of the heterointerfaces are considered. The spectrally and
time-resolved resonant Rayleigh scattering (RRS) spectra calculated for the interface model are in remarkable agreement with both the time-resolved and
spectrally resolved measurements, indicating that the main features of the exciton disorder potential are captured by the model. Calculating the RRS using
exciton eigenstates and microscopically calculated dephasing rates due to radiative decay and phonon scattering instead of the often used time-propagation
model predicts the observed spectrally dependent homogeneous linewidth and the non-exponential RRS dynamics at long times. The calculated localized exciton
states in the monolayer islands including their phonon scattering and their dipole moments are also relevant for comparison with experiments on individual
excitonic states performed using high spatial resolution \cite{WuPRL99, GuestScience01,LangbeinPRL05}. Our result indicates that the present disorder model
contains all the important features of semiconductor heterointerfaces fabricated by growth-interrupted molecular beam epitaxy, and can substantially improve
our understanding in a large variety of experimental situations.

\acknowledgments{The sample was grown by K. Leosson and J.Riis\,Jensen at the
III-V Nanolab, a joint laboratory between Research Center COM and the Niels
Bohr Institute, Copenhagen University. The authors thank K.\,Leosson and
G.\,Kocherscheidt for contributions to the experiments. V. S. acknowledges
financial support from the Swiss National Science Foundation, through project
No. 620-066060.}


\end{document}